\documentclass[a4paper,11pt]{article}
\usepackage{graphicx}
\usepackage{amsmath,amsfonts,amssymb}
\usepackage[bbgreekl]{mathbbol}
\usepackage[margin=1in]{geometry}
\usepackage{jheppub}
\usepackage{amssymb}
\usepackage{braket}
\usepackage{todonotes}
\usepackage{subcaption}
\usepackage{comment}
\usepackage{makecell}
\usepackage{cleveref}

\setcellgapes{2pt}

\DeclareSymbolFontAlphabet{\mathbbm}{bbold}
\DeclareSymbolFontAlphabet{\mathbb}{AMSb}

\title{Where is tree-level heterotic string theory?}

\author[a]{Mathieu Boisvert,}
\author[b]{Waltraut Knop,}
\author[a]{and Leonardo Rastelli}

\affiliation[a]{C. N. Yang Institute for Theoretical Physics, Stony Brook University,\newline Stony Brook, NY 11794-3840, U.S.A.}

\affiliation[b]{Deutsches Elektronen-Synchrotron DESY, Notkestr. 85, 22607 Hamburg, Germany}

\preprint{DESY-26-072}

\abstract{We continue the tree-level S-matrix bootstrap program for quantum gravity, now in ten-dimensional theories with half-maximal supersymmetry. This setting includes the heterotic string and allows us to test whether the string-like structures found in the maximally supersymmetric bootstrap persist with less supersymmetry and in non-planar gauge sectors. Imposing analyticity, crossing symmetry, unitarity,  and Regge boundedness, we constraint weakly coupled UV completions of supergravity and super Yang-Mills. In the gravitational sector, much of the boundary of the allowed region is explained by extremal amplitudes supported on a single linear Regge trajectory, or by convex combinations of such amplitudes. In the gluon sector, the non-planar problem reveals a tension between representation-channel positivity and the trace decomposition of the EFT data, obstructing a direct normalization by $G$ or $g_{\rm YM}$; a coupled gluon/graviton bootstrap  may be necessary to directly constrain the relative strength of gauge and gravitational interactions. Overall, our results support the emergence of linear Regge trajectories as a robust feature of the tree-level quantum-gravity bootstrap.}

\begin{document}

\maketitle

\section{Introduction and summary}

A sharp formulation of the problem of quantum gravity is to ask what becomes of the gravitational S-matrix at high energies. At low energies, gravity is universal and is described by Einstein's theory. At sufficiently high energies, however, the gravitational interaction must be completed by new degrees of freedom. String theory provides the canonical example of such a completion already at weak coupling: an infinite tower of massive higher-spin states appears at the string scale, and tree-level amplitudes exhibit remarkable conspiracies accommodating analyticity, unitarity, crossing symmetry and Regge behavior.

The question we would like to address is how much of this structure is forced by general principles alone. In other words, can one characterize the space of possible weakly coupled tree-level completions of gravity without assuming a worldsheet description or a string spectrum as microscopic input? This is the perspective of the modern S-matrix bootstrap (see~e.g.~\cite{Kruczenski:2022lot, deRham:2022hpx} for reviews). One parametrizes theory space by low-energy EFT data, together with some spectral information about the first massive states, and imposes the consequences of unitarity and causality. The hope is that the resulting allowed regions in theory space are small enough, and structured enough, to reveal the organizing principles of quantum gravity.

In previous work \cite{Albert:2024yap},  continuing the analysis started in~\cite{Caron-Huot:2021rmr}, we  pursued this program for maximally supersymmetric tree-level amplitudes in ten dimensions. The assumptions were deliberately minimal. We considered scattering of massless supergravitons and supergluons, imposed a mass gap, tree-level unitarity, crossing symmetry, and Regge boundedness strong enough to require the reggeization of the graviton or gluon. Supersymmetry enters not as a sacred principle but as a simplifying assumption: it relates the scattering of gravitons and gluons to a much simpler auxiliary amplitude and improves its high-energy behavior, rendering the dispersive analysis much simpler.\footnote{See the {\it tour the force} papers~\cite{Caron-Huot:2022ugt, Caron-Huot:2022jli,Pasiecznik:2025eqc} for constraints on higher-derivative corrections to (non-supersymmetric) Einstein gravity.} Following the blueprint of~\cite{Arkani-Hamed:2020blm, Bellazzini:2020cot, Tolley:2020gtv, Caron-Huot:2020cmc,Caron-Huot:2021rmr}, these assumptions lead to sum rules expressing low-energy Wilson coefficients in terms of positive spectral densities, supplemented by null constraints encoding crossing. The resulting convex optimization problem gives rigorous exclusion plots for suitably normalized EFT data.

The main surprise of \cite{Albert:2024yap} was not that string theory obeyed the bounds, but rather where it appeared within them. Type II and Type I string theory lived safely in the interior of the allowed regions. The extremal boundaries were instead controlled by more economical amplitudes, whose extracted spectra appeared to consist of a single Regge trajectory. In the supergraviton problem, the distinguished extremal solution had the same leading Regge slope as the Type II string, while in the supergluon problem (see also~\cite{Berman:2023jys,Berman:2024wyt} for a very similar analysis) the boundary was generated by a continuous family of single-trajectory solutions with varying slope. Thus a very stringy structure emerged from the bootstrap, even though the full string amplitude, with its daughter trajectories, was not itself extremal.

This paper asks whether this phenomenon is special to maximal supersymmetry. We study the analogous problem for ten-dimensional theories with $\mathcal{N}=(1,0)$ supersymmetry. This is the natural setting for the heterotic string. It is also the next simplest arena in which to test the generality of the picture just described. Reducing supersymmetry makes the bootstrap problem technically richer and physically more revealing. In the gravitational sector, new low-energy interactions are allowed, including the $R^2$ interaction familiar from the Green--Schwarz mechanism, and the Newton constant can no longer be disentangled from the EFT data in quite the same way as in the maximally supersymmetric case. In the gauge sector, one can study gluon scattering without imposing the planar limit. The heterotic string then becomes a natural target, with both single- and double-trace structures present, and with positivity most naturally formulated in representation channels rather than in the trace basis.

Let us also place our approach in the broader landscape of recent work. The non-perturbative S-matrix bootstrap program  aims to carve out the space of unitary, crossing-symmetric gravitational amplitudes at finite coupling \cite{Guerrieri:2021ivu,Guerrieri:2022sod}. This is conceptually close in spirit, but technically rather different from our weakly coupled tree-level setup, where unitarity reduces to positivity of spectral densities. More closely related to our perturbative regime is the work of~\cite{Elvang:2026pmc}, who have emphasized the power of higher-point amplitudes: in planar maximally supersymmetric gauge theory, supersymmetry, $R$-symmetry and factorization of five- and six-point EFT amplitudes impose nonlinear constraints on four-point Wilson coefficients, and, when combined with positivity, appear to single out the open-string Veneziano amplitude~\cite{Elvang:2026pmc}.

A distinct line of work has been pursued by the authors of~\cite{Arkani-Hamed:2023jwn,Cheung:2022mkw,Cheung:2024uhn,Cheung:2025nhw,Cheung:2025tbr}, who identify striking structural features of string amplitudes, such as strong high-energy softness, level truncation or the pattern of zeros governing the cancellation of higher-spin exchanges, and elevate them to axioms from which the uniqueness of the amplitude, and even its spectrum, follows.  Our emphasis is complementary. We wish to assume only the fundamental principles, and ask which of these features are forced upon us, so that stringy structure appears as a consequence of consistency rather than as an input.

The present work should also be viewed as a technical step toward more general mixed-system bootstrap problems. The non-planar gluon amplitude leads naturally to a matrix-valued bootstrap problem: positivity is diagonal in irreducible representation channels, while the low-energy EFT data are organized in a trace basis, and the two descriptions are related by nontrivial linear transformations. We also develop more systematic diagnostics for interpreting the geometry of the numerical exclusion plots, distinguishing boundary components associated with isolated low-spin states, towers at the cutoff or at infinity, and genuine Regge-like extremal solutions. These tools are not specific to the heterotic example, and should be useful in future bootstraps involving several coupled amplitudes. We now summarize our setup and  main results.

\subsection*{Half-maximal supergraviton scattering}
We first consider $2\to2$ scattering of massless $\mathcal{N}=(1,0)$ supergravity multiplets in ten dimensions. Since a fully ten-dimensional on-shell superspace formalism for this multiplet is not available, we use the standard four-dimensional on-shell superspace after choosing the external kinematics to lie in a four-dimensional subspace. Supersymmetry then reduces a certain set of component amplitudes  to a crossing-symmetric auxiliary scalar amplitude $\mathcal{M}(s,u)$. This amplitude obeys a dispersive representation with positive spectral density, together with an infinite set of null constraints following from crossing. The crucial novelty, compared to the maximally supersymmetric case, is the presence of a constant term $g_0$ in the low-energy expansion of $\mathcal{M}(s,u)$. This coefficient is allowed by half-maximal supersymmetry and is nonzero in the heterotic string. It also changes the structure of the sum rules. In particular, $g_0$ does not arise from an independent positive sum rule once the amplitude is normalized by the Newton constant~$G$. Rather, it mixes with the graviton-exchange contribution. As a consequence, the allowed region is unbounded in the positive $g_0$ direction: one may add purely non-gravitational supersymmetric scalar amplitudes and increase $g_0$ without changing the normalization by~$G$.

Despite this degeneracy, the bootstrap gives nontrivial bounds. We derive exclusion plots in the plane of the first two EFT coefficients normalized by  the Newton constant $G$ and the cutoff. Both the heterotic and Type II string amplitudes lie inside the allowed region. The boundary has a rich structure. Part of it can be understood from amplitudes already familiar in related scalar bootstrap problems, such as an infinite spin tower at the cutoff. Once appropriate spectral assumptions are imposed to remove such unphysical towers, new extremal solutions become visible.

The main result in the gravitational sector is that the boundary is generated by two families of amplitudes whose spectra organize into single Regge trajectories. One family has fixed slope, equal to the leading string Regge slope, while the intercept is continuously shifted relative to the graviton pole. The second family keeps the mass of the first massive state fixed while the slope varies, interpolating between a trajectory with string-like slope and a limiting configuration in which an infinite spin tower accumulates at the cutoff. These structures are closely analogous to the extremal families found in the maximally supersymmetric Type II and Type I analyses, but they appear here in a more elaborate form because half-maximal supersymmetry allows a larger space of EFT deformations.

It would be naive to identify these extremal amplitudes with complete theories of quantum gravity. Indeed, amplitudes supported on a single Regge trajectory are expected to be too sparse to satisfy all consistency conditions~\cite{Eckner:2024pqt,Eckner:2025kve}. Plausibly, they arise as limiting solutions of healthy families of amplitudes with a richer spectrum of states. String theory itself sits in the interior,  because its daughter trajectories  provide positive deformations of these extremal single-trajectory solutions.

\subsection*{Non-planar supergluon scattering}

We next turn to $2\to2$ scattering of massless vector multiplets. Famously, anomaly cancellation and supersymmetry~\cite{AlvarezGaume:1983ig, Green:1984sg, Adams:2010zy}  restrict the allowed gauge groups to $SO(32)$ and $E_8 \times E_8$. We focus on the $SO(32)$ case -- a very similar analysis could be carried out for $E_8 \times E_8$. Unlike in the open-string setup of \cite{Albert:2024yap,Berman:2023jys,Berman:2024wyt}, we do not impose a planar limit. The full amplitude contains both single- and double-trace structures, corresponding respectively to gauge and gravitational exchange in the low-energy theory. However, these two amplitudes are not independent from the point of view of unitarity. Positivity is diagonal in the representation-channel basis for the product of two adjoint representations of $SO(N)$, and the transformation from the trace basis to this representation basis mixes the single- and double-trace sectors.

This mixing leads to the principal obstruction of the gluon analysis. The natural sum rules associated with the Yang--Mills coupling and with Newton's constant are matrix-valued in representation space. The corresponding kernels are not positive definite: their entries have mixed signs across the six representation channels. Consequently, neither $g_{\rm YM}^2$ nor $G$ can be used directly as a positive normalization in the standard convex bootstrap setup. One can otherwise increase the quantity to be bounded by adding spectral weight in one channel and compensate the normalization by adding weight in another channel with the opposite sign.

We therefore introduce an auxiliary positive normalization, chosen at the lowest mass dimension for which the corresponding heavy average converges for both the heterotic and Type I string amplitudes. With this normalization we obtain bounds on the first accessible Wilson coefficients in the single- and double-trace sectors. The resulting allowed region is simple: its extremal points are supported in individual representation channels, up to states at infinity needed to satisfy null constraints. Unlike in the gravitational analysis, we do not see a clean emergence of Regge trajectories. This is likely not a deep physical statement about heterotic gluon scattering, but rather a limitation of the present convex setup. At the subtraction level accessible with the auxiliary normalization, isolated low-spin exchanges remain allowed and the Regge constraints are too weak to force the appearance of infinite towers.

\bigskip
\noindent
The remainder of the paper is organized as follows. In Section~\ref{sec:gravity} we treat graviton scattering, deriving the sum rules and presenting the bounds and their extremal solutions. In Section~\ref{sec:gluons} we turn to gluon scattering, establish the obstruction to the standard procedure, and present the bounds obtainable with the auxiliary normalization. We conclude in Section~\ref{sec:discussion} with a discussion of the overall picture emerging from our results and of possible future directions. Two appendices contain further technical material. Appendix~\ref{app:wavepacket} introduces a modified wavepacket basis for smearing the graviton pole. Appendix~\ref{app:extraSR} collects additional sum rules for gluon scattering.

\section{Graviton scattering}\label{sec:gravity}

In this section, we study $2\to2$ scattering of massless $\mathcal{N}=(1,0)$ supergravity multiplets in $D=10$. Our goal is to derive nontrivial bounds on the low-energy effective field theory (EFT) coefficients governing higher-derivative corrections to Einstein gravity. Supersymmetry allows us to encode the relevant four-point function in terms of an auxiliary scalar amplitude. In contrast to the maximally supersymmetric case \cite{Albert:2024yap},  the Newton constant cannot be disentangled from certain EFT coefficients, which substantially alters the structure of the resulting bounds. 

We begin by deriving the supersymmetric representation of the amplitude and establishing its general analytic structure. We then use these properties to construct sum rules and null constraints, which form the basis for the numerical bootstrap analysis and the resulting bounds on the EFT coefficients.

\subsection{Preliminaries}

We begin by reviewing how supersymmetry constrains the amplitude and then analyze the additional constraints imposed by analyticity, unitarity, and Regge behavior. Finally, we comment on how these general features are realized in heterotic and Type II string theory.

\subsubsection{Super Ward identities}
The $2 \to 2$ scattering amplitude of massless $\mathcal{N} = (1,0)$ gravity multiplets in $D = 10$ is constrained by supersymmetry through super Ward identities, see \cite{Elvang:2013cua} for a pedagogical review and \cite{Albert:2024yap} for similar cases with varying amounts of supersymmetry and different multiplets. 

Note that the $D=10$, $\mathcal{N} = (1,0)$ SUSY algebra does not admit a known on-shell superspace. We thus follow the same strategy as \cite{Albert:2024yap} and work in the 4d on-shell superspace. Since we are only concerned with four-point amplitudes, Lorentz transformations allow us to choose a frame in which the external kinematics lie in a four-dimensional subspace, so this reduction from 10d to 4d entails no loss of generality. Our strategy will thus be to express the constraints coming from the super Ward identities in this 4d language, where we will solve them to extract a nice auxiliary amplitude that will be amenable to the bootstrap.

Having outlined the strategy, we now introduce the on-shell formalism we will use. The 16 supercharges of $D=10$, $\mathcal{N} = (1,0)$ supersymmetry arrange into $D = 4$ complex Weyl spinors $Q^I_{\alpha}, \bar{Q}_{J\dot{\beta}}$ where $\alpha, \dot{\beta}$ are 4d Weyl indices and $I,J$ are fundamental $SU(4)_R$ indices. We define the usual spinor-helicity variables $p_{\alpha\dot{\beta}} = |p]_\alpha \bra{p}_{\dot{\beta}}$. In this language, on the massless multiplet, the supercharges satisfy
\begin{equation}
    \{Q^I_{\alpha}, \bar{Q}_{J\dot{\beta}} \} = |p]_\alpha \bra{p}_{\dot{\beta}} \delta^I_J\, .
\end{equation}
Upon introducing fermionic coordinates $\eta_I$, an on-shell superspace is given by 
\begin{equation}
    Q^I_\alpha = |p]_\alpha \frac{\partial}{\partial \eta_I}, \quad \bar{Q}_{J\dot{\beta}} = \bra{p}_{\dot{\beta}} \eta_J\, .
\end{equation}
Having obtained this $D = 4$ on-shell realization of the algebra, we now determine how the $D = 10$ fields manifest themselves in this language. Upon reduction to four dimensions, the $D=10$ gravity multiplet splits into 6 copies of the $D=4$ abelian vector multiplet, $\Phi^i_V$, $i = 1,\cdots,6$, plus a $D=4$ gravity multiplet, $\Phi_G$. The $D = 4$ vector multiplet is obtained by acting with $\bar{Q}$ on a Clifford vacuum of helicity ($-1$) and can be represented by the following superfield
\begin{equation}
    \Phi^i_V(p, \eta) = \phi^i_- + \eta_I \psi^{iI} + \frac{1}{2!}\eta_I \eta_J\phi^{iIJ} + \frac{1}{3!}\eta_I\eta_J\eta_K\psi^{iIJK} + \eta_I\eta_J\eta_K\eta_L \phi^i_+\,,
\end{equation}
where $\phi_{\pm}$ are the components of the gauge field with helicity $\pm1$, $\phi^{IJ}$ are scalar fields in the $\textbf{6}$ of $SU(4)_R$ and $\psi^{IJK}$ are the fermionic partners. The $D=4$ gravity multiplet is obtained by acting with $\bar{Q}$ on two distinct Clifford vacua of helicity ($-2$) and ($0$) respectively. Two vacua are required to ensure CPT invariance, with CPT exchanging the two sets of states. It admits the following superfield expansion:
\begin{align}
\begin{split}
    \Phi_G(p, \eta) &= \Psi_G + \tilde{\Psi}_G \\ &=(g_{-2} + ... + \eta^1\eta^2\eta^3\eta^4\phi) + (\bar{\phi} + ... + \eta^1\eta^2\eta^3\eta^4 g_{+2})\, ,
    \end{split}
\end{align}
where $g_{\pm2}$ are the components of the graviton with $h = \pm 2$, $\phi$ and $\bar{\phi}$ make up the axio-dilaton and the ellipses contain the remaining polarization states that complete the on-shell $D=4$ gravity multiplet.

The $D = 10$ gravity multiplet is reducible when viewed from the perspective of $D=~4$, $\mathcal{N} = 4$ supersymmetry. As a result, unlike in the maximally supersymmetric case, the $2 \to 2$ gravity amplitude cannot be expressed in terms of a single auxiliary amplitude. Instead, one must consider several distinct amplitudes involving the superfields $\Phi_V, \Phi_G$. More explicitly, decomposing $\Phi_G$ into its irreducible components $\Psi_G, \tilde{\Psi}_G$, we obtain the following non-vanishing superamplitudes\footnote{Other combinations vanish by $R$-charge counting, see \cite{Albert:2024yap} for closely related examples.}
\begin{equation}\label{eq:superamps}
    \langle\Phi^i_V\Phi^j_V\Phi^k_V\Phi^l_V \rangle\, ,\quad \langle\Phi^i_V\Phi^j_V\Psi_G\tilde{\Psi}_G \rangle\, ,\quad \langle\Phi_G\Phi_G\tilde{\Psi}_G\tilde{\Psi}_G \rangle\, .
\end{equation}
In contrast with the maximally supersymmetric case, there are now multiple independent component amplitudes. We focus on a configuration of external states the defines a convenient auxiliary amplitude. Specifically, we take four identical copies of the $D=4$ vector multiplet, corresponding to the first superamplitude in \eqref{eq:superamps}, with $i=j=k=l$. This choice ensures full crossing symmetry and is the only one with this property. The super Ward identities then admit a unique solution of the form
\begin{equation}
    \mathcal{A}_{4V}(p_i, \eta_i) = \delta^{10}(p_1+p_2+p_3+p_4)\,\delta^8(\bar{Q})\,\mathcal{M}(s,u)\, ,
\end{equation}
where the fermionic delta function is 
\begin{equation}
    \delta^8(\bar{Q}) = \frac{1}{2^4}\prod\bar{Q}_{I\dot{\beta}}\bar{Q}^{\dot{\beta}}_I = \frac{1}{2^4}\prod^4_{I=1} \sum^4_{i,j=1} \langle ij\rangle \eta_{iI} \eta_{jI}\, ,
\end{equation}
with $(i,j)$ labeling the external legs, and $p_i$ the corresponding momenta. Because the external superfields are identical, $\mathcal{A}_{4V}(p_i, \eta_i)$ is fully crossing symmetric, a property which the reduced amplitude $\mathcal{M}(s,u)$ inherits:
\begin{equation}\label{eq:crossingsymmetry}
    \mathcal{M}(s,u) = \mathcal{M}(u,s) = \mathcal{M}(t,u)\, ,
\end{equation}
where $s,t,u$ are the Mandelstam variables with $t=-s-u$.

We now extract the component of the superamplitude corresponding to the scattering process $\phi_-\phi_- \to \phi_+\phi_+$. This is obtained by projecting onto the term proportional to $\prod_{I}\eta_{3I}\eta_{4I}$ in $\delta^8(\bar{Q})$, yielding 
\begin{equation}\label{eq:factors}
    A(\phi_-\phi_- \to \phi_+\phi_+) = s^2 \mathcal{M}(s,u)\, ,
\end{equation}
where $A$ denotes the corresponding component amplitude. The function $\mathcal{M}(s,u)$ will serve as the object of our bootstrap analysis in subsequent sections. In $D=10$ language, $A$ corresponds to a particular choice of graviton and $B$-field polarizations.

\subsubsection{Analyticity, unitarity, and low energy expansion}

Having obtained the reduced amplitude $\mathcal{M}(s,u)$ through the super Ward identities, we now study its properties under basic consistency conditions such as analyticity, unitarity, Regge boundedness and consistency with low-energy EFT expansion. We keep $D$ arbitrary in this section, specializing to $D=10$ when presenting numerical bounds.

We will only consider tree-level extensions of $\mathcal{N}=(1,0)$ supergravity in $D=10$. Thus, the amplitude $\mathcal{M}(s,u)$ is analytic in the complex $s$ plane at fixed $u$, except at points on the real axis corresponding to tree-level exchanges of on-shell states. Since \eqref{eq:factors} relates the reduced amplitude to the physical amplitude, a spin-$J$ exchange in $\mathcal{M}$ generally corresponds to a linear combination of physical exchanges with spins up to $J+2$, depending on the external states used to extract the reduced amplitude.

For a $2\rightarrow 2$ process, unitarity amounts to requiring that on-shell couplings are real or, for our purposes, that their squares are non-negative. Strictly speaking, this condition applies to the physical amplitudes rather than directly to the reduced amplitude $\mathcal{M}$. Nevertheless, positivity of the partial-wave expansion
\begin{equation}\label{eq:partialwave}
    \text{Im}[\mathcal{M}(s,u)] = s^{(4-D)/2}\!\sum_{J=\text{even}}\! n_J^{(D)}\rho_J(s)\,\mathcal{P}_J\!\left(\mbox{$1+\frac{2u}{s}$}\right)
\end{equation}
is clearly a necessary condition, as the auxiliary spectral density is related by a factor of $s^2$ to the physical spectral density of  $A(\phi_-\phi_- \to \phi_+\phi_+)$. We expect it to be sufficient for all amplitudes in the supersymmetric orbit of
$A(\phi_-\phi_- \to \phi_+\phi_+)$.  
One indication is obtained by considering crossed amplitudes for which the supersymmetric prefactor is proportional to $t^2$ rather than $s^2$. In that channel, since products of Gegenbauer polynomials decompose into finite positive linear combinations of Gegenbauer polynomials, positivity of the coefficients in the reduced amplitude is manifestly sufficient for unitarity of the physical amplitude.  It would be interesting to investigate whether an analysis of the complete 10d supergraviton scattering process might lead to stronger constraints.

We therefore impose the condition
\begin{equation}
\rho_J(s) \geq 0 \quad \text{for} \quad s>0,, u<0, .
\end{equation}
Our conventions for the Gegenbauer polynomials and the partial wave normalization are as follows:
\begin{align}
    \begin{split}
        \mathcal{P}_J(x) &= {}_2F_1\!\left(\mbox{$-J,J+D-3,\frac{D-2}{2},\frac{1-x}{2}$}\right)\, ,
        \\[4pt]
        n_J^{(D)} &= \frac{2^D\pi^\frac{D-2}{2}}{\Gamma\!\left(\mbox{$\frac{D-2}{2}$}\right)}(J+1)_{D-4}\,(2J+D-3)\, ,
    \end{split}
\end{align} 
where $(x)_n = \frac{\Gamma(x+n)}{\Gamma(x)}$ is the Pochhammer symbol. 
The sum in \eqref{eq:partialwave} is restricted to even spins since the amplitude is fully crossing symmetric. This can easily be seen from the behavior of the Gegenbauer polynomials under the exchange $t \leftrightarrow u$:
\begin{equation}
    \mathcal{P}_J\!\left(\mbox{$1+\frac{2u}{s}$}\right) \, \longleftrightarrow\, \mathcal{P}_J\!\left(\mbox{$1 + \frac{2(-s-u)}{s}$}\right) = (-1)^J \,\mathcal{P}_J\!\left(\mbox{$1 + \frac{2u}{s}$}\right)\, ,
\end{equation}
leading to the vanishing of odd-spin contributions.

Our final assumption on the full amplitude is that of strict Regge boundedness, meaning that the physical amplitude grows strictly slower than $O(s^2)$ in the limit of large $s$ at fixed negative $u$. We will sometimes refer to an amplitude with Regge behavior $O(s^J)$ as exhibiting spin-$J$ Regge behavior.  The 
expectation of Regge behavior $J \leq 2$ is supported by causality arguments \cite{Camanho:2014apa}, connections with the chaos bounds in holographic theories \cite{Chandorkar:2021viw}, and more general considerations of gravitational Regge behavior \cite{Chowdhury:2019kaq,Haring:2022cyf}. Strict $J <2$ Regge boundedness is a stronger condition.
As discussed in \cite{Albert:2024yap}, this assumption is equivalent to the Reggeization of the graviton: rather than remaining an isolated spin-two state, the graviton is assumed to lie on a Regge trajectory extending to higher spin.
It is also what
ensures that gravity participates non-trivially in the bootstrap problem, since otherwise the Newton constant would remain unconstrained by the dispersion relations.

Because this condition applies to the physical amplitude $A$, the auxiliary amplitude grows strictly slower than $O(s^0)$ in the Regge limit, thanks to \eqref{eq:factors}:
\begin{equation}\label{eq:gravityRegge}
    \lim_{|s|\to\infty} \mathcal{M}(s,u) \to 0 \quad \text{for fixed} \quad u<0\, .
\end{equation}
This improved Regge behavior is a direct consequence of supersymmetry, via the super Ward identities. In the maximally supersymmetric case, the four graviton scattering amplitude has Regge behavior improved by four units. As we will see later, the improvement of the Regge behavior due to supersymmetry will allow us to obtain additional sum rules.

Up to this point, all consistency conditions have been imposed on the full amplitude at arbitrary energies. We now introduce additional assumptions that define the low-energy regime. We assume a mass gap $M$ between the massless states and the first excited state(s), so that $M$ sets the scale below which the dynamics are well described by an effective field theory. This implies that, within a finite neighborhood of the origin, the amplitude has no poles other than those associated with the gravity multiplet, so that the amplitude admits a general crossing-symmetric low-energy expansion of the form
\begin{equation}\label{eq:MIRgrav}
    \mathcal{M}_{\text{IR}}(s,u) = 8\pi G\left(\frac{s}{tu} + \frac{t}{us} + \frac{u}{st} \right) + g_0 + g_2\,(s^2+t^2+u^2) + g_3 \,(stu)+\,\cdots\, .
\end{equation}
The first term corresponds to the exchange of massless states, i.e., the graviton and its superpartners, in the $D=10$ gravity multiplet. Note that its behavior in the Regge limit is $O(s^0)$, meaning in isolation it is marginally excluded by our assumption of strict Regge boundedness. We then parametrize the low-energy EFT by the most general expansion consistent with crossing symmetry, with a priori arbitrary coefficients $g_i$ multiplying higher-derivative interactions. For example, the coefficient $g_0$ arises from an $R^2$ interaction and is related to the Green--Schwarz anomaly cancellation mechanism, while $g_2$ captures the effect of an $R^4$ interaction, and so on. Our goal is to place bounds on these Wilson coefficients, normalized relative to the gravity exchange. While all terms included above are allowed by $D=10$, $\mathcal{N}=(1,0)$ supersymmetry, not all are compatible with maximal supersymmetry, as we will see explicitly when considering the Type II amplitude in the next section.

\subsubsection{String theory amplitudes}

There are two known UV complete amplitudes that satisfy all the assumptions spelled out above, arising from the heterotic and Type II string theories. Although Type II string theory exhibits maximal supersymmetry, $\mathcal{N}=(2,0)$ or $(1,1)$, the corresponding amplitude can be consistently interpreted within the half-maximal framework. 

The corresponding reduced amplitudes, defined through \eqref{eq:factors}, are crossing symmetric and exhibit Regge growth of the form $s^{\alpha' u/4}$, consistent with \eqref{eq:gravityRegge}. While a direct proof of unitarity in terms of positivity of residues is not available, it follows from the worldsheet no-ghost theorem \cite{Goddard:1972,Aoki:1990}. In addition, they admit a low-energy expansion of the form \eqref{eq:MIRgrav}.

In the heterotic string, the tree-level amplitude for $2 \to 2$ scattering of states in the gravity multiplet takes the form \cite{KAWAI19861}
\begin{equation}\label{eq:hetamplitude}
    A_{4g}(\{p^{(i)},\epsilon^{(i)}\}) =\left(\prod_{i}\epsilon^{(i)}_{\mu_i\nu_i}\right)K_{\text{bos}}^{\mu_1\mu_2\mu_3\mu_4}K_{\text{ss}}^{\nu_1\nu_2\nu_3\nu_4}\mathcal{M}_{VS}(s,u)\, ,
\end{equation}
where the $\epsilon_{\mu\nu}$ are polarization tensors, $K_{\text{bos}}$ and $K_{\text{ss}}$ are the kinematic factors of the bosonic string and superstring respectively, whose expressions can be found in \cite{KAWAI19861, Basu:2017nhs} and $\mathcal{M}_{VS}$ is the Virasoro--Shapiro amplitude, given by\footnote{We use the convention of \cite{green1987superstring1}.} 
\begin{equation}\label{eq:VSamplitude}
    \mathcal{M}_{\text{VS}}(s,u) = -\frac{\Gamma\left(\mbox{$-\frac{\alpha' s}{4}$}\right) \Gamma\left(\mbox{$-\frac{\alpha' t}{4}$}\right) \Gamma\left(\mbox{$-\frac{\alpha' u}{4}$}\right)}{\Gamma\left(\mbox{$1+\frac{\alpha' s}{4}$}\right) \Gamma\left(\mbox{$1+\frac{\alpha' t}{4}$}\right) \Gamma\left(\mbox{$1+\frac{\alpha' u}{4}$}\right)}\, ,
\end{equation}
where $\alpha'$ is the inverse string tension. Decomposing the polarization tensors $\epsilon^{(i)}_{\mu\nu}$ into symmetric, antisymmetric and trace part, yields amplitudes for gravitons, Kalb--Ramond $B$-fields and dilatons, respectively. Evaluating \eqref{eq:hetamplitude} for the relevant polarizations yields the reduced amplitude
\begin{equation}
    \mathcal{M}_{\text{Het}}(s,u) =\left(\frac{16\alpha'^2(s^2+t^2+u^2)+24\alpha'^3(stu)-\alpha'^4(s^4+t^4+u^4)}{8(4+\alpha's)(4+\alpha't)(4+\alpha'u)}\right)\mathcal{M}_{VS}(s,u)\, .
\end{equation}
The factor of $s^2$ required by the super Ward identities \eqref{eq:factors} is contained in the kinematic factors in \eqref{eq:hetamplitude} and has been stripped off. Expanding at low energies, we extract
\begin{equation}\label{eq:Hetvalues}
    \text{Heterotic:}\qquad8\pi G = \frac{1}{\alpha'}\, , \quad g_0 = 6\, , \quad  g_2 = (2+ 2\zeta(3))\, \alpha'^2 \approx 4.4 \,\alpha'^2\, .
\end{equation}
In the Type II case, the corresponding amplitude is\footnote{This is the relevant amplitude when considering the Type II theory as a theory with half maximal SUSY.} \cite{Berman:2024eid}
\begin{equation}
    \mathcal{M}_{\text{Type II}}(s,u) = \alpha'^2 (s^2+t^2+u^2)\mathcal{M}_{VS}(s,u)\, ,
\end{equation}
leading to
\begin{equation}\label{eq:VSvalues}
    \text{Type II:}\qquad8\pi G = \frac{64}{\alpha'}\, , \quad g_0 = 0\, , \quad g_2 = 2\zeta(3)\alpha'^2\approx 2.4\,\alpha'^2\, .
\end{equation}
Here $g_0=0$ reflects the absence of an $R^2$ interaction, which is incompatible with maximal supersymmetry.

As a reminder, the spectrum of these amplitudes organizes into Regge trajectories, with masses $m^2 = \frac{4}{\alpha'} n$, $n \in \mathbb{N}$. The leading trajectory passes through the graviton exchange, which in the reduced amplitude appears as an effective spin-zero exchange. The first massive state on this trajectory has spin $J=2$.

\subsection{Sum rules and null constraints}

Having established the general constraints from supersymmetry, analyticity, and unitarity, we now derive sum rules that encode these properties as quantitative constraints on the amplitude, following the method and notation of \cite{Caron-Huot:2020cmc,Caron-Huot:2021rmr,Albert:2024yap}.

We begin with the usual contour integral of the amplitude along an arc at infinity in the complex $s$-plane at fixed $u<0$. As discussed in the previous section, Regge boundedness ensures that for a suitable number ($k$) of subtractions, this contribution vanishes. This defines a family of sum rules $\mathcal{C}_k$:
\begin{equation}
    \mathcal{C}_k:\quad\frac{1}{2\pi i}\oint_\infty \frac{\mathrm{d}s}{s}\frac{\mathcal{M}(s,u)}{[s(s+u)]^{k/2}}  = 0 \quad \text{for} \quad k=0,2,4,\ldots\, .
\end{equation}
Deforming the contour and collapsing it onto the singularities along the real $s$-axis splits the result into a low-energy (IR) and a high-energy (UV) contribution:
\begin{align}
    \underset{s=0,-u}{\mathrm{Res}}\left[\frac{\mathcal{M}_{\text{IR}}(s,u)}{s[s(s+u)]^{k/2}}\right]
    &= \int\displaylimits_{M^2}^\infty \frac{\mathrm{d}m^2}{\pi}\left(\frac{1}{m^2}+\frac{1}{m^2+u}\right)\frac{\mathrm{Im}[\mathcal{M}(m^2,u)]}{[m^2(m^2+u)]^{k/2}} 
    \\
    &\equiv \left<\frac{2m^2+u}{m^2+u}\frac{\mathcal{P}_J\!\left(\mbox{$1+\frac{2u}{m^2}$}\right)}{[m^2(m^2+u)]^{k/2}}\right>\, .
\end{align}
In the first line, we used crossing symmetry to relate the two cuts. In the second line, we inserted the partial-wave expansion, \eqref{eq:partialwave}, for the UV amplitude and introduced the shorthand
\begin{equation}\label{eq:heavyavg}
    \left<(\cdots)\right>\equiv\frac{1}{\pi}\sum_{J \text{ even}}  n^{(D)}_J\int\displaylimits_{M^2}^\infty\frac{\mathrm{d}m^2}{m^2} m^{4-D}\rho_J(m^2)\,(\cdots)\, ,
\end{equation}
which we will refer to as the heavy average. By unitarity, the spectral density is positive, and therefore the measure defining this heavy average is positive.

Inserting the EFT expansion of the amplitude, \eqref{eq:MIRgrav}, the first three sum rules are
\begin{equation}\label{eq:Csumrules}
    \begin{alignedat}{2}
        \mathcal{C}_0: &\quad & -\frac{2}{u}8\pi G + g_0 + 2 u^2 g_2 + \cdots &= \left< \frac{(2 m^2+u)\mathcal{P}_J\!\left(\mbox{$1+\frac{2u}{m^2}$}\right)}{m^2+u} \right>\, , 
        \\
        \mathcal{C}_2: &\quad & 2 g_2 - u g_3 + 8 u^2 g_4 + \cdots &= \left< \frac{(2 m^2+u)\mathcal{P}_J\!\left(\mbox{$1+\frac{2u}{m^2}$}\right)}{m^2 (m^2+u)^2} \right>\, , 
        \\
        \mathcal{C}_4: &\quad & 4 g_4 - 2 u g_5 + 24 u^2 g_6 + \cdots &= \left< \frac{(2 m^2+u)\mathcal{P}_J\!\left(\mbox{$1+\frac{2u}{m^2}$}\right)}{m^4 (m^2+u)^3} \right>\, .
    \end{alignedat}
\end{equation}
For $k\geq 2$, expanding around the forward limit ($u\to 0$) allows us to isolate individual EFT coefficients, for example
\begin{equation}\label{eq:g2SumRule}
    g_2 = \left< \frac{1}{m^4}\right>, \quad g_3 = \left<\frac{3-\frac{1}{2}J(7+J)}{m^6}\right>\, .
\end{equation}
Positivity of the heavy-average measure immediately implies that $g_2\geq 0$.

As was first discussed in \cite{Caron-Huot:2021rmr}, the graviton contribution in $\mathcal{C}_0$ cannot be isolated in this manner due to the presence of the $1/u$ pole. We construct an improved sum rule by subtracting appropriate combinations of higher-$k$ sum rules (which do not contain the pole) evaluated in the forward limit:
\begin{equation}
    \mathcal{C}_0^{\text{improved}}=\mathcal{C}_0-\sum_{n=2}^\infty u^n (\mathcal{C}_n|_{u=0})\, .
\end{equation}
This yields
\begin{equation}\label{eq:SRgravG0}
    \mathcal{C}_0^{\text{improved}}:\quad-\frac{2}{u}8\pi G + g_0=\left<\frac{(2 m^2+u)\mathcal{P}_J\!\left(\mbox{$1+\frac{2u}{m^2}$}\right)}{m^2+u}-\frac{2 u^2}{m^4-u^2}\right>\, .
\end{equation}
The key structural point is that gravity and the first EFT coefficient $g_0$ appear in a coupled way. Unlike the maximally supersymmetric case, gravity cannot be completely isolated. It is intrinsically tied to $g_0$. A similar coupled sum rule was first identified for massless scalar scattering coupled to gravity in \cite{Caron-Huot:2021rmr}.

Because of the $1/u$ pole, the forward limit of this improved sum rule is not well defined. Following \cite{Caron-Huot:2021rmr,Albert:2024yap}, we instead smear both sides against wavepackets $f_k(p)=p^{k+1/2}$ with $k=1,2,3,\ldots$, and $u=-p^2$:
\begin{equation}\label{eq:Gsmeared}
    \int\displaylimits_0^{M} \mathrm{d}p \,f_k(p)\left(\frac{8\pi G}{p^2} + g_0\right)  =\left<\int\displaylimits_0^{M} \mathrm{d}p \,f_k(p)\left(\frac{(2 m^2+u)\mathcal{P}_J\!\left(\mbox{$1+\frac{2u}{m^2}$}\right)}{m^2+u}-\frac{2 u^2}{m^4-u^2}\right)\right>\, .
\end{equation}
We therefore obtain a family of sum rules that express the low-energy data in terms of heavy averages over the unknown UV spectrum. Our objective is to use these relations to bound the low-energy observables, normalized relative to gravity.

We now exploit crossing symmetry more fully. While the low-energy ansatz $\mathcal{M}_{\text{IR}}$ is manifestly symmetric under $(s\leftrightarrow u \leftrightarrow t)$, the UV representation in terms of partial waves is not. By repeating the contour deformation with a crossing-antisymmetric kernel, we derive additional constraints on the spectral density. Since these constraints depend only on UV data and vanish for the IR contribution, they are referred to as {\it null constraints} \cite{Caron-Huot:2020cmc, Albert:2023jtd}. They follow from
\begin{equation}\label{eq:NCint}
    \oint_0 \frac{\mathrm{d}u}{2\pi i}\oint_\infty \frac{\mathrm{d}s}{2\pi i}\frac{1}{s u}\left(\frac{\mathcal{M}(s,u)}{s^{n-\ell}u^\ell}-\frac{\mathcal{M}(u,s)}{u^{n-\ell}s^\ell}\right)=0,\quad \text{for}\,\,n\geq \ell \geq 0\, .
\end{equation}
The resulting constraints take the form $\left<\mathcal{X}_{n,\ell}\right>=0$, where
\begin{equation}\label{eq:NCgravity}
    \mathcal X_{n,\ell} = \mathop{\mathrm{Res}}_{u = 0}\left[\frac{1}{u}\left(\left(\frac{1}{m^{2(n-\ell+1)}u^\ell}-\frac{1}{u^{n-\ell}m^{2\ell+2}}\right)-(m^2\rightarrow -m^2-u)\right)m^2\,\mathcal{P}_J\!\left(\mbox{$1+\tfrac{2u}{m^2}$}\right)\right]\, .
\end{equation}
The set of independent constraints is given by $n=4,5,\ldots$, and $l=1,2,\ldots,\left[\mbox{$\frac{n-1}{3}$}\right]$. The null constraints begin at one higher subtraction level, consistent with Regge behavior in the presence of the graviton pole.

The sum rules relate the EFT coefficients to heavy averages over the UV spectrum, while the null constraints impose additional restrictions on the spectral density. Positivity of the heavy-average measure together with the linearity of these constraints implies that the space of allowed spectral data is convex. This allows us to derive rigorous bounds on EFT coefficients, normalized relative to the Newton's constant and the cutoff.

The resulting optimization problem can be recast as a semidefinite program. Define
\begin{equation}
    \Vec v_{\text{HE}}\equiv\left(-G_1(m^2,J),\,\ldots\,,-G_{k_\text{max}}(m^2,J)\,,\mathcal X_{4,1}(m^2,J)\,,\,\ldots\,,\mathcal X_{n_{\text{max}},\left[\frac{n_{\text{max}}-1}{3}\right]}(m^2,J)\right)\, ,
\end{equation}
where $G_k$ denotes the integrand appearing inside the heavy average in the smeared sum rule \eqref{eq:Gsmeared} (with smearing functions $f_k$) and we have included a finite set of null constraints. We further define
\begin{align}
    \begin{split}
        \Vec v_{\text{obj}} &\equiv\left(2 M^{1/2},\,\ldots\,, \tfrac{M^{k_\text{max}-1/2}}{k_\text{max}-1/2}, 0\, ,\,\ldots\,,0\right)\, , 
        \\
        \Vec v_{\text{norm}} &\equiv\left(\tfrac{2}{5}M^{5/2},\,\ldots\,, \tfrac{M^{k_\text{max}+3/2}}{k_\text{max}+3/2}, 0\, ,\,\ldots\,,0\right)\, ,
    \end{split}
\end{align}
which encode the coefficients of $8\pi G$ and $g_0$, respectively, on the left-hand side of the smeared sum rule. With these definitions, the bootstrap equation takes the form
\begin{equation}\label{eq:g0BootstrapEq}
    8\pi G\,\Vec v_{\text{obj}}+ g_0\,\Vec v_{\text{norm}}+ \left<\Vec v_{\text{HE}}(m^2,J)\right>=0\, .
\end{equation}
The semidefinite program is thus to find a vector $\vec{\alpha}$ that maximizes $\vec{\alpha} \cdot \vec{v}_{\text{obj}}$, subject to the constraints $\vec{\alpha} \cdot \vec{v}_{\text{norm}} = \pm1$ and $\vec{\alpha} \cdot \vec{v}_{\text{HE}}(m^2,J) \geq 0$ for all even $J$ and $m \geq M$. If such a vector exists, it provides a bound on the coefficient (upper or lower depending on the sign of $\Vec\alpha\cdot\Vec v_{\text{norm}}$). These bounds can only improve as additional wavepackets and null constraints are included, and are therefore rigorous.

In principle, the positivity condition must hold for all $m \geq M$ and all even $J$, but in practice the problem is rendered computationally tractable by introducing a suitable truncation in both variables and focusing on certain limits. Since this truncation amounts to working on a discrete mesh in $(m,J)$ space, one must also check for possible negative regions between sampled points and refine the mesh accordingly. The details of this truncation procedure can be found in \cite{Albert:2024yap}. The only modification in our setup is that the form of the smeared gravity sum rules is no longer polynomial in $m$ (in \cite{Albert:2024yap} it is polynomial of degree $2J$), so we discretize in $m$, even for low spins. Once this truncation is applied, the resulting semidefinite program can be input into \texttt{SDPB} \cite{Simmons-Duffin:2015qma}.

If we want to incorporate assumptions on the spectrum, there are two complementary approaches. To bound an EFT coefficient, one can restrict the positivity condition $\vec{\alpha}\cdot\vec{v}_{\text{HE}}(m^2,J)\geq 0$ to the subset of states that are allowed by the assumed spectrum and proceed as before. A more explicit implementation, introduced in \cite{Albert:2023seb}, is to include a massive resonance directly in the spectral density,
\begin{equation}\label{eq:rhoShift}
    n_{J}^{(D)}\rho_{J}(m^2)\,\longrightarrow\, \lambda_{\phi}^2\,\pi\,\delta_{J,J_\phi}\delta(m^2-m_{\phi}^2)\,m_\phi^{D-2} + n_{J}^{(D)}\widetilde{\rho}_{J}(m^2)\, ,
\end{equation}
where $\lambda_\phi$ denotes the three-point coupling between two gravitons and the exchanged state of mass $m_\phi$ and (effective) spin $J_\phi$. This then modifies the heavy averages and consequently the bootstrap equation as discussed in \cite{Albert:2023seb,Albert:2024yap}, allowing us to bound the on-shell coupling directly. Here we fix the overall scale by $m_\phi$ and treat the cutoff ratio $M/m_\phi \geq 1$ as an external parameter.

Before presenting the resulting bounds, it is useful to highlight an important structural feature of the optimization problem. The linearity of the sum rules and null constraints implies that the space of allowed spectral densities is convex. Consequently, if two spectral densities satisfy all constraints, then any positive linear combination of them is also an allowed solution.

As emphasized above, Newton's constant and $g_0$ appear jointly in the coupled sum rule \eqref{eq:SRgravG0}. Unlike in the maximally supersymmetric case, the gravitational contribution cannot be isolated. A similar coupled sum rule was first identified in the context of massless scalar scattering coupled to gravity in \cite{Caron-Huot:2021rmr}. An immediate consequence is that $g_0$ does not admit an upper bound.

To see this, consider solutions for which $G=0$. In that case, the sum rules imply that $g_0$ is positive. By convexity, any such solution can be added to a solution of the gravitational system with an arbitrary positive coefficient. Since our observables are normalized relative to $G$, this operation leaves the normalization unchanged while increasing $g_0$. It follows that $g_0$ can be made arbitrarily large.

On the other hand, $g_0$ does admit a nontrivial lower bound. In particular, it can even become slightly negative, as observed in \cite{Caron-Huot:2021rmr}, since the additional presence of gravity in the sum rule can compensate for a slight negative contribution from $g_0$. In units of $G$ and $M$, with $k_{\text{max}}=8$ and $n_{\text{max}}=17$, we obtain the lower bound
\begin{equation}
    \tilde{g}_0 \equiv \frac{g_0 M^2}{8\pi G}\geq -7.08\, .
\end{equation}
Our immediate goal is to determine the exclusion region in the $(\tilde{g}_0$,$\tilde{g}_2)$ plane, with $\tilde{g}_2 \equiv \frac{g_2 M^6}{8\pi G}$. 
To do so, we must first characterize the non-gravitational solutions that can be consistently added to the gravitational system. This system was studied in \cite{Berman:2024eid} and describes supersymmetric scalar scattering. We first review the relevant results in the next section before returning to the gravitational analysis.

\subsection{Bounds}

In this section, we derive bounds on the lowest-dimension EFT coefficients. We begin by reviewing the non-gravitational analysis of supersymmetric scalar scattering presented in \cite{Berman:2024eid}, and then turn to bounds normalized relative to gravity.

\subsubsection{Recap: Bounds without gravity}

\begin{figure}[htbp]
    \centering
    \begin{subfigure}{.49\textwidth}
        \centering
        \includegraphics[width=\linewidth]{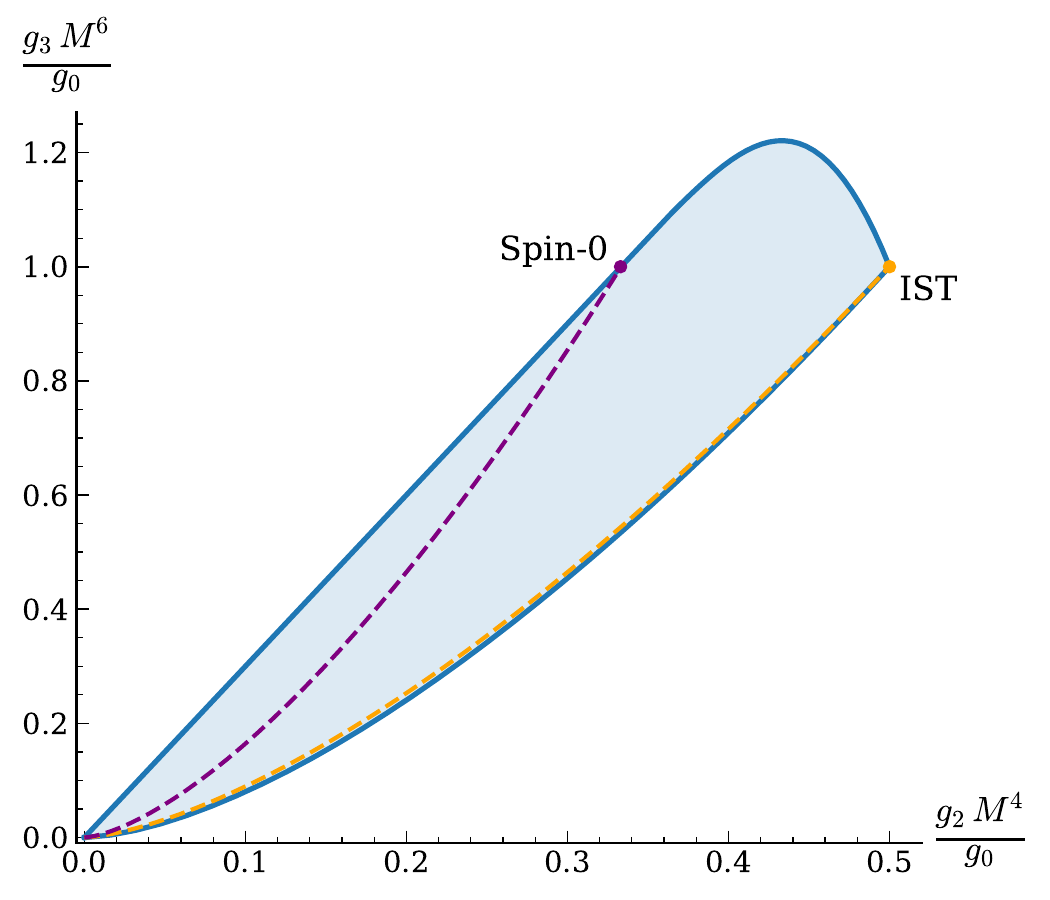}
        \caption{No spectral assumptions}
        \label{fig:g2g3nospec}
    \end{subfigure}
    \begin{subfigure}{.49\textwidth}
        \centering
        \includegraphics[width=\linewidth]{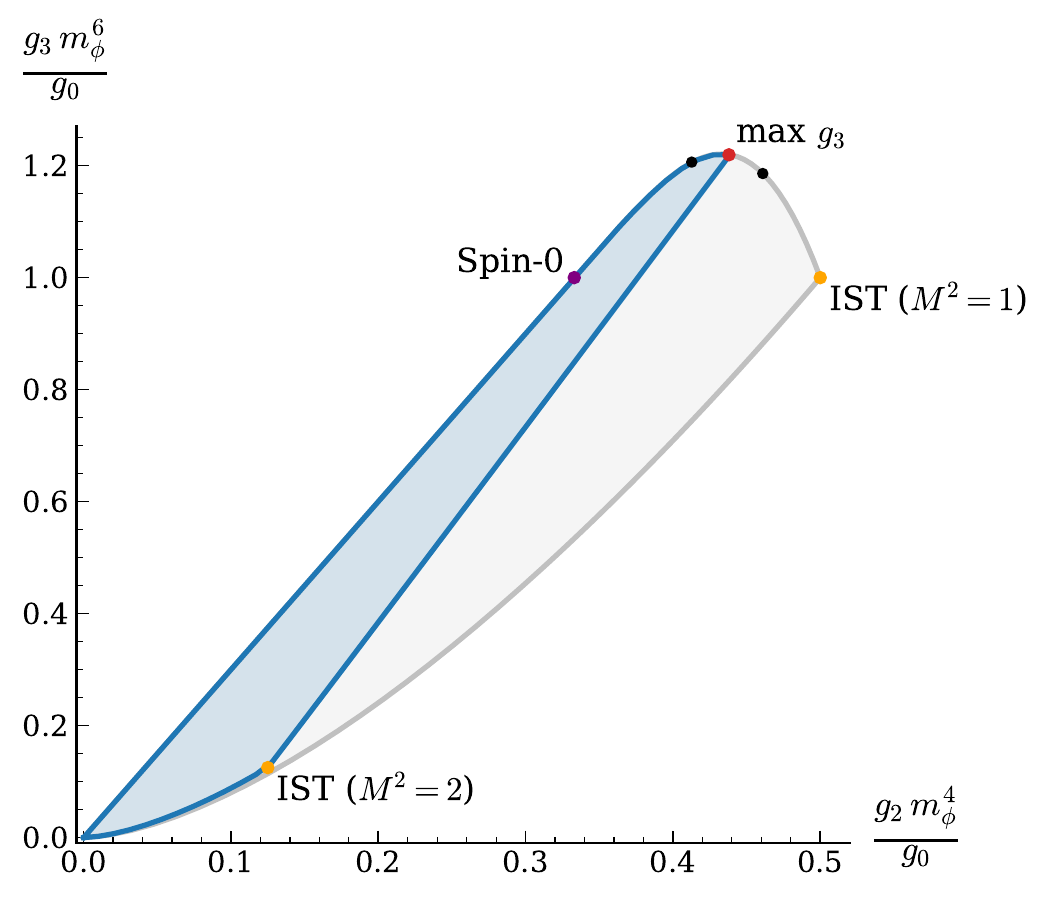}
        \caption{With spectral assumptions}
        \label{fig:g2g3wspec}
    \end{subfigure}
    \caption{Exclusion plots for $g_2$ and $g_3$ in units of $g_0$ and either the cutoff $M$ in \ref{fig:g2g3nospec} or the mass of the first resonance $m_\phi$ in \ref{fig:g2g3wspec}. The allowed region is shaded in blue. In the absence of spectral assumptions (left), the boundary is controlled by two characteristic families of amplitudes: the infinite spin tower \eqref{eq:towerofstates} (orange, IST) and the scalar exchange \eqref{eq:spin0exch} (purple, spin-zero). The marked point corresponds to the case where the mass scale coincides with the cutoff, while the dotted curve shows the continuation as the mass is taken above the cutoff. Imposing a spectral assumption (right) by restricting the first resonances at mass $m_\phi$ to spins $0$ and $2$ and introducing a gap up to the cutoff $M^2=2m_\phi^2$, shrinks the allowed region. Two additional kinks appear: one associated with the shifted infinite tower (orange), and another (red) corresponding to the point that maximizes $g_3$. Spectra at the two black points as well as at the red point are shown in figure~\ref{fig:noGravspec}. Both plots were generated with $n_{\text{max}}=15$.}
    \label{fig:g2g3plot}
\end{figure}

In the absence of gravity, the forward limit of the sum rules \eqref{eq:Csumrules} yields\footnote{In contrast to the gravitational case, no smearing is required here, so the positivity condition $\vec{\alpha}~\cdot~\vec{v}_{\text{HE}}(m^2,J)~\geq ~0$ can be taken to be polynomial in $m^2$, with convergence ensured by selecting a suitably dense set of spins $J$, following \cite{Caron-Huot:2020cmc}.}
\begin{equation}
    g_0 = \left< 2\right>\, ,\qquad 
    g_2 = \left< \frac{1}{m^4}\right>\, ,\qquad 
    g_3 = \left<\frac{3-\frac{1}{2}J(7+J)}{m^6}\right>\, .
\end{equation}
In this case, one obtains an additional level of null constraints arising from the lowest-subtraction sum rule. As a result, the full set of independent null constraints in \eqref{eq:NCgravity} is specified by $n=1,2,\cdots$ and $\ell=0,1,\cdots,\left[\frac{n-1}{3}\right]$. We can therefore bound $g_2$ and $g_3$, normalized by $g_0$ and the cutoff $M$. The resulting exclusion region is shown in figure~\ref{fig:g2g3nospec}.

We briefly summarize the key features of these bounds, referring to \cite{Berman:2024eid} for a more detailed analysis. The boundary of the allowed region is controlled by three characteristic structures. First, the maximum of $g_2 M^4/g_0$ is saturated by an amplitude with an infinite spin tower (IST) at the cutoff,
\begin{equation}\label{eq:towerofstates}
    \mathcal{M}_{\text{IST}}= -\frac{M^6}{(M^2-s)(M^2-u)(M^2-t)}\, .
\end{equation}
Varying the mass scale of this tower relative to the cutoff traces out the lower boundary of $g_3 M^6/g_0$ (orange curve in figure~\ref{fig:g2g3nospec}).

Second, a linear segment of the boundary is generated by a single scalar exchange at the cutoff,
\begin{equation}\label{eq:spin0exch}
    \mathcal{M}_{\text{spin-zero}}=\frac{M^2}{M^2-s}+\frac{M^2}{M^2-u}+\frac{M^2}{M^2-t}\, .
\end{equation}
Finally, a curved segment interpolates between these two amplitudes. As shown in \cite{Berman:2024eid}, the amplitudes along this portion are characterized by a scalar resonance at the cutoff followed by a gap to higher mass states, with the gap increasing as one moves from the infinite tower towards the scalar exchange.

We can further restrict the allowed region by imposing spectral assumptions that exclude infinite towers. In particular, suppose that the first mass threshold at $m_\phi$ contains only spins 0 and 2, and that there is a gap above this first resonance such that the cutoff satisfies $M^2 = 2 m_\phi^2$. Under these assumptions, the allowed region shrinks, as shown in figure~\ref{fig:g2g3wspec}. The maximal value of $g_2 m_\phi^4/g_0$ decreases to $\approx0.44$ and is attained at the point that globally maximizes $g_3 M^6/g_0$.

\begin{figure}[htbp]
    \centering
    \begin{subfigure}{0.32\textwidth}
        \centering
        \includegraphics[width=\linewidth]{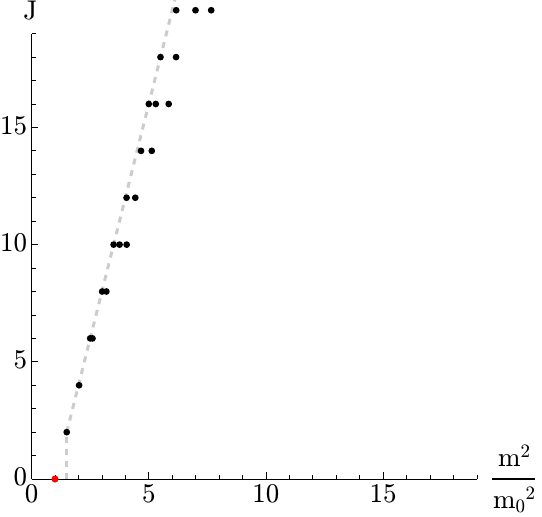}
        \caption{$\frac{M^2}{m_\phi^2} = \tfrac{3}{2}$}
    \end{subfigure}
    \hfill
    \begin{subfigure}{0.32\textwidth}
        \centering
        \includegraphics[width=\linewidth]{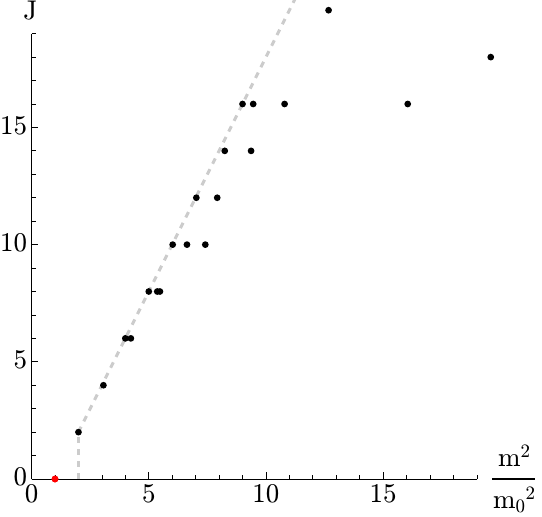}
        \caption{$\frac{M^2}{m_\phi^2} = 2$}
    \end{subfigure}
    \hfill
    \begin{subfigure}{0.32\textwidth}
        \centering
        \includegraphics[width=\linewidth]{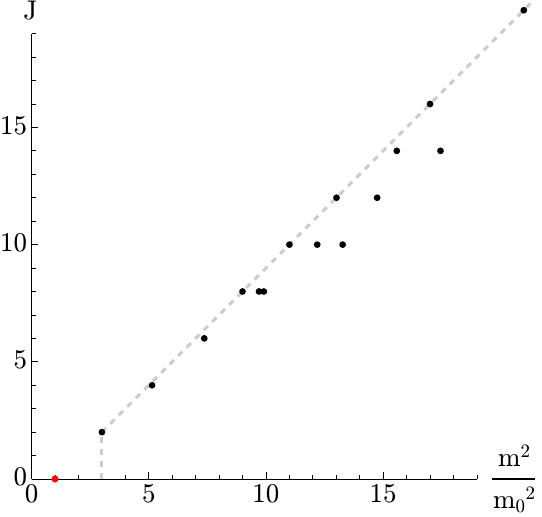}
        \caption{$\frac{M^2}{m_\phi^2} = 3$}
    \end{subfigure}
    \caption{Chew--Frautschi plots generated with \texttt{SDPB} by maximizing $g_2 m_\phi^4/g_0$, allowing for a scalar exchange at $m_\phi$ and varying the cutoff $M$. The three spectra correspond to representative points along the curved boundary marked by the two black points and and the red point in figure~\ref{fig:g2g3wspec}. The red point in the Chew--Frautschi plots indicates the imposed scalar resonance, while the black points are extracted from \texttt{SDPB}. The dashed gray line denotes the maximal spin-dependent lower bound that can be imposed without modifying the bound on $g_2 m_\phi^4/g_0$. All plots were produced with $n_{\text{max}}=19$.}
    \label{fig:noGravspec}
\end{figure}

To gain further insight into the amplitudes saturating the curved boundary, we extract the corresponding spectra from \texttt{SDPB}. As discussed in \cite{Albert:2024yap}, the raw spectra typically contain spurious poles. These can be systematically removed by imposing additional spectral constraints that leave the bound unchanged. The resulting cleaned spectra for representative values $M^2/m_\phi^2 = \tfrac{3}{2}, 2, 3$ are shown in figure~\ref{fig:noGravspec}.

These spectra correspond to the marked points in figure~\ref{fig:g2g3wspec}. They suggest that the extremal solutions form a family of single Regge trajectories with varying slope, passing through a scalar first resonance. This behavior closely parallels what was observed in the open-string analysis of \cite{Albert:2024yap}. Note that we impose only spin-dependent lower bounds on the masses, without restricting the mass from above. Whether the higher-mass portion of the spectrum can be further simplified, or instead reflects intrinsic structure, remains an open question.

For later use, we summarize the key features. Without spectral assumptions, the ratio $g_2 M^4/g_0$ is maximized by the infinite spin tower, yielding $g_2 M^4/g_0 = 1/2$. Imposing a gap and restricting the first resonance to spins $0$ and $2$ excludes this solution. The new extremal point is instead described by a single Regge trajectory, maximizing $g_3 M^6/g_0$, and satisfying $g_2 m_\phi^4/g_0 \approx 0.44$.

\subsubsection{Bounds with Gravity}\label{sec:gravitybounds}

\begin{figure}[ht]
    \centering
    \includegraphics[width=0.85\textwidth]{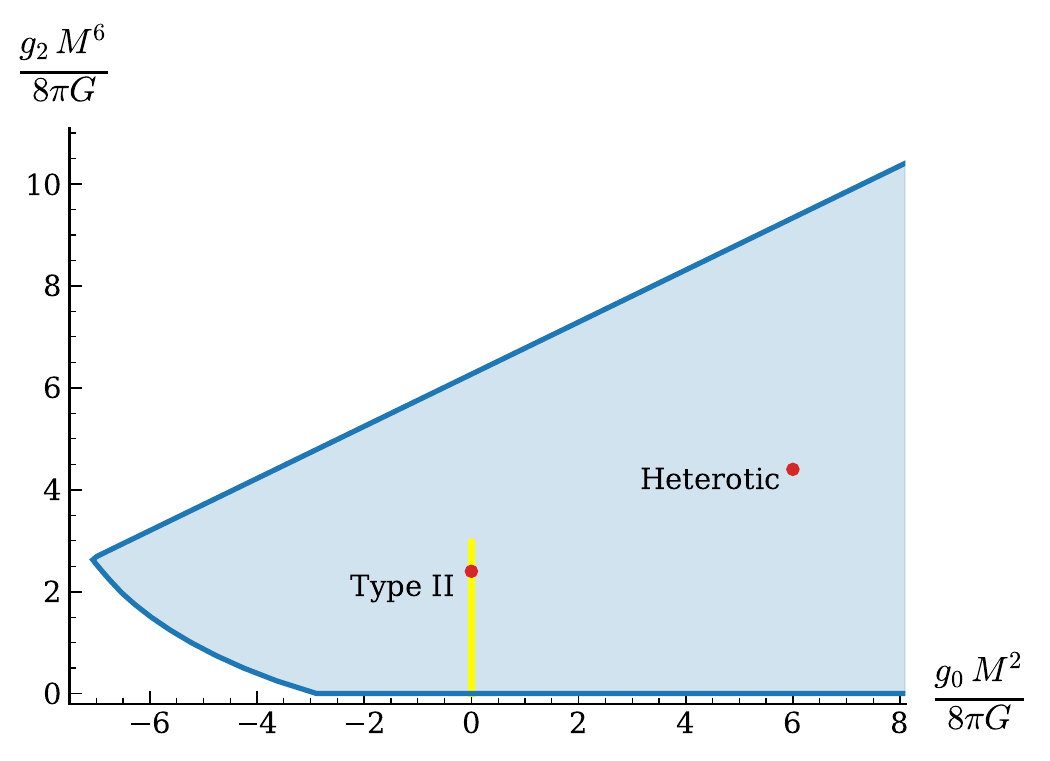}
    \caption{Exclusion plot in the $(g_0,g_2)$ plane, in units of $G$ and $M$. The blue shaded region denotes the allowed parameter space. The bounds are obtained numerically using $k_{\text{max}}=7$ and $n_{\text{max}}=15$ (7 wavepackets and 30 null constraints). The red dots correspond to the Virasoro--Shapiro amplitude \eqref{eq:VSvalues} (Type II) and the heterotic string amplitude \eqref{eq:Hetvalues}, both with $M^2\alpha'/4=1$. The yellow line shows the bounds with maximal supersymmetry from \cite{Albert:2024yap}.}
    \label{fig:HGr_g0g2wST}
\end{figure}

We now turn to the full system including gravity. Normalizing all EFT coefficients by the Newton's constant, $G$, and the cutoff, $M$, we obtain the exclusion region in the $(\tilde g_0,\tilde g_2)$ plane shown in figure~\ref{fig:HGr_g0g2wST}. Consistent amplitudes lie within the shaded region, which is unbounded as $\tilde g_0,\tilde g_2 \to \infty$. However, at fixed $\tilde g_0$, the coefficient $\tilde g_2$ is bounded both from above and below.

\begin{figure}[ht]
    \centering
    \includegraphics[width=0.8\textwidth]{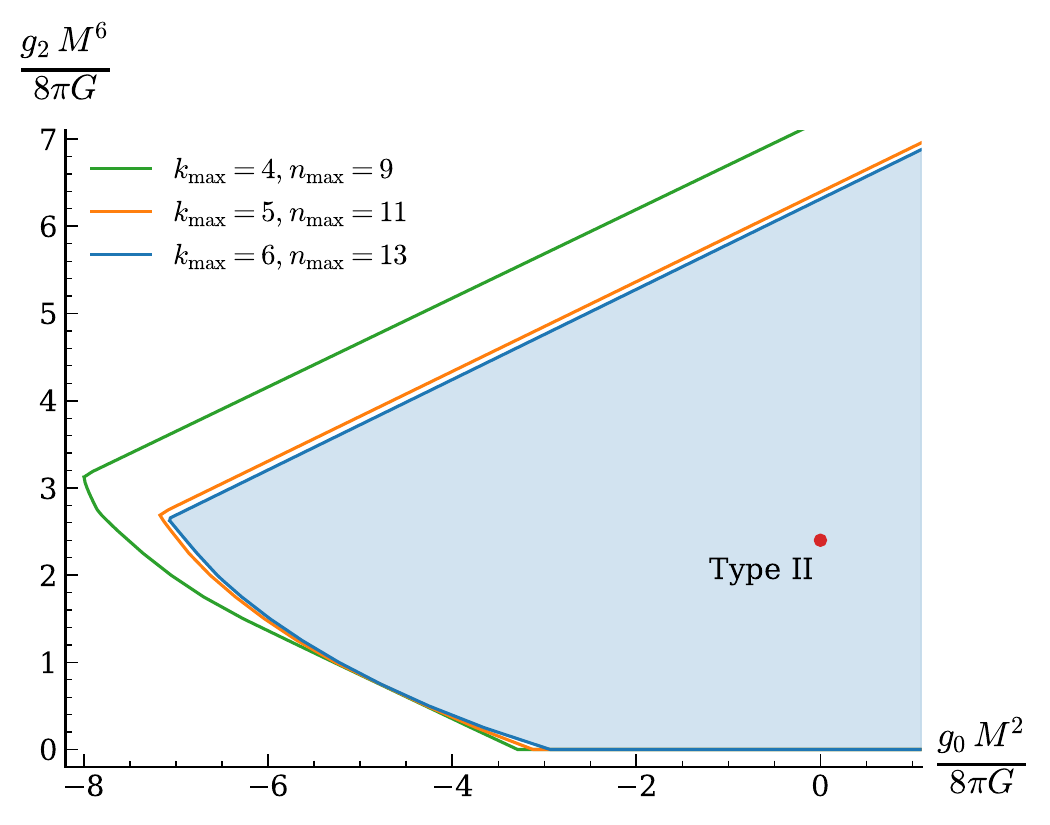}
    \caption{Convergence of the exclusion region with increasing numbers of wavepackets and null constraints. At this scale, changes in the bounds become visually indistinguishable for larger values except for the kink at $\tilde g_2$ which converge more slowly towards $\tilde g_0 \approx -2.7$.}
    \label{fig:HGr_g0g2nmax}
\end{figure}

We assess convergence by increasing the number of null constraints and wavepackets, as shown in figure~\ref{fig:HGr_g0g2nmax}. For $k_{\text{max}} \geq 6$ and $n_{\text{max}}\geq 13$, the bounds are stable at the resolution of the plot. We therefore have chosen to fix $k_{\text{max}}=7$ and $n_{\text{max}}=15$ unless otherwise stated.

As expected, the string theory amplitudes \eqref{eq:Hetvalues} and \eqref{eq:VSvalues} lie within the allowed region. Comparing with the maximally supersymmetric case \cite{Albert:2024yap}, we note that maximal supersymmetry enforces $g_0=0$ and removes all terms of the form $(stu)^n$ in the EFT expansion \eqref{eq:MIRgrav}. This leads to significantly stronger bounds along the $\tilde g_2$-axis, as indicated by the yellow line in figure~\ref{fig:HGr_g0g2wST}.

We now describe the main structural features of the allowed region. The point minimizing $\tilde g_0$ is located at $(\tilde g_0,\tilde g_2)\approx(-7.08,2.65)$. From this point, the upper boundary extends linearly towards large $\tilde g_0$ with slope $1/2$. The $\tilde g_0$-minimizing point connects to the $\tilde g_2=0$ axis through a curved segment. The $g_2$ sum rule \eqref{eq:g2SumRule} implies that approaching $g_2=0$ requires pushing all resonances to parametrically large mass, $m\to\infty$.

A linear boundary can be associated with a non-gravitational solution that maximizes the ratio $g_2 M^4/g_0$. Convexity allows such a solution to be added with arbitrary weight, generating a ray in the ($\tilde g_2$, $\tilde g_0$) plane whose slope is precisely this ratio. In the present case, the slope is 1/2, which precisely coincides with the maximal value of $g_2 M^4/g_0$ in the non-gravitational problem and is attained by the infinite spin tower amplitude \eqref{eq:towerofstates}. One might therefore conclude that the linear boundary is generated simply by adding the non-gravitational infinite spin tower solution. However, as we will see below, this interpretation is incomplete. 

In the numerical analysis, convergence around $\tilde g_2=0$ is noticeably slower than at nonzero $\tilde g_2$, see figure \ref{fig:HGr_g0g2nmax}. It can be improved by augmenting the wavepacket basis with a localized contribution at the cutoff scale. This is discussed in App. \ref{app:wavepacket}. However, we do not make systematic use of this refinement in the remainder of the analysis, since it does not significantly affect the results at nonzero $\tilde g_2$ and complicates the large-spin and small-mass approximations required for increasing the cutoff.

\subsubsection{Understanding extremal points}

To better understand the boundary, we follow \cite{Albert:2024yap} and impose a spectral assumption. In this reference, the first resonance was taken to be a scalar separated by a gap from the cutoff, excluding infinite spin tower spectra and isolating amplitudes with finitely many lowest-lying states.

In the present case, the softer Regge behavior requires allowing both spin-zero and spint-two states at the first mass $m_\phi$. However, any scalar contribution can be removed without violating unitarity, thereby decreasing $\tilde g_0$. Extremal amplitudes in this system therefore do not contain scalar resonances.

\begin{figure}[ht]
    \centering
    \includegraphics[width=0.85\textwidth]{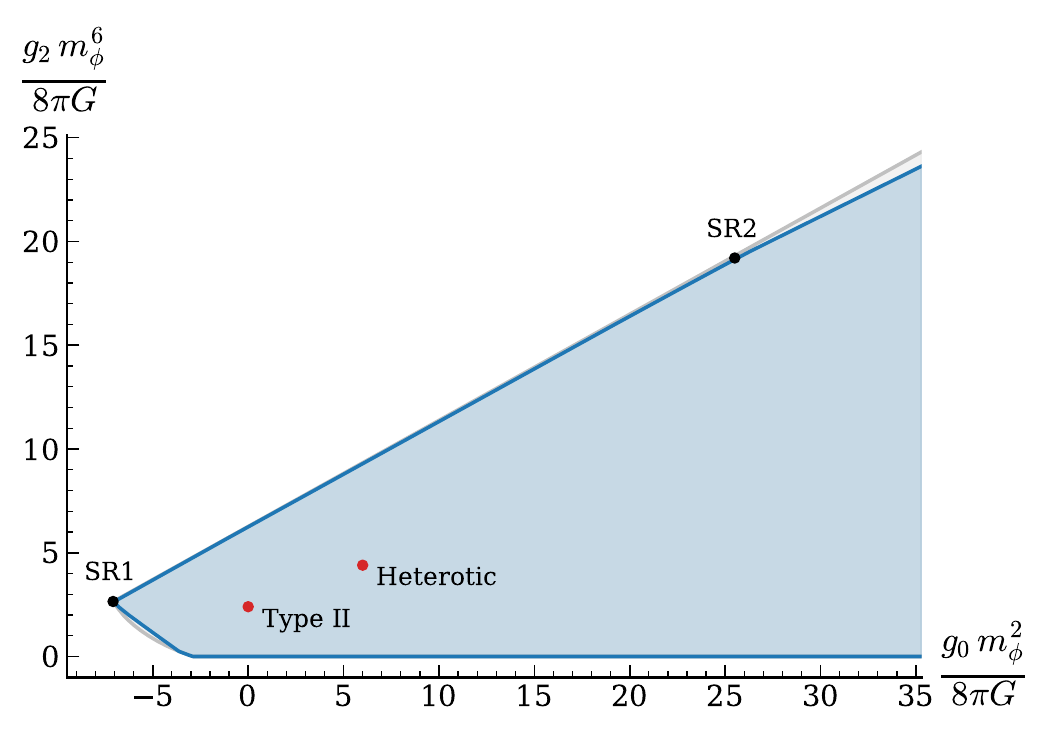}
    \caption{Exclusion plot in the $(g_0,g_2)$ plane, in units of $G$ and the mass $m_\phi$ of the first resonance, with a cutoff $M^2 = 2 m_\phi^2$. The allowed region is shaded in blue, while the gray region shows the corresponding bounds obtained without spectral assumptions for comparison. The imposed spectral assumptions are satisfied by string theory, whose values are indicated by the red dots. The effect of the spectral assumptions on the bounds is mild and localized to two regions. At negative $\tilde g_0$, the curved lower bound on $\tilde g_2$ is partially removed. The kink at the global minimum of $\tilde g_0$ remains and is marked by the black dot labeled SR1. On the upper boundary, the linear behavior with slope $1/2$ persists over a wide range. However, at large $\tilde g_0$ (around $\tilde g_0 \approx 25.5$), a kink develops and the slope decreases to $\approx 0.44$. This point is marked by the black dot labeled SR2.}
    \label{fig:HGr_g0g2full}
\end{figure}

Imposing this assumption and normalizing by $m_\phi$, we recompute the exclusion region for various cutoffs $M$. For $M>m_\phi$, the tower-of-states amplitude responsible for the linear upper bound is excluded. The resulting exclusion region for $M^2=2m_\phi^2$ is shown in figure~\ref{fig:HGr_g0g2full}.

Several features are immediately apparent. The lower bound $\tilde g_0=0$ is unchanged, as it is saturated by amplitudes with all resonances at infinite mass and hence is unaffected by the spectral assumption. The curved segment at negative $\tilde g_0$ is partially removed, while the $\tilde g_0$-minimizing point (SR1) persists. Notably, a segment of the linear upper bound with slope $1/2$ remains, despite the absence of tower-of-states amplitudes. At larger $\tilde g_0$, a kink develops (SR2), beyond which the slope decreases to $\approx 0.44$, matching the value associated with the $g_3/g_0$-maximizing amplitude in the non-gravitational analysis. Since these two changes to the bounds are separated by an order of magnitude and are hard to resolve in the same plot we proceed by analyzing them individually. 

\begin{figure}[ht]
    \centering
    \includegraphics[width=0.8\textwidth]{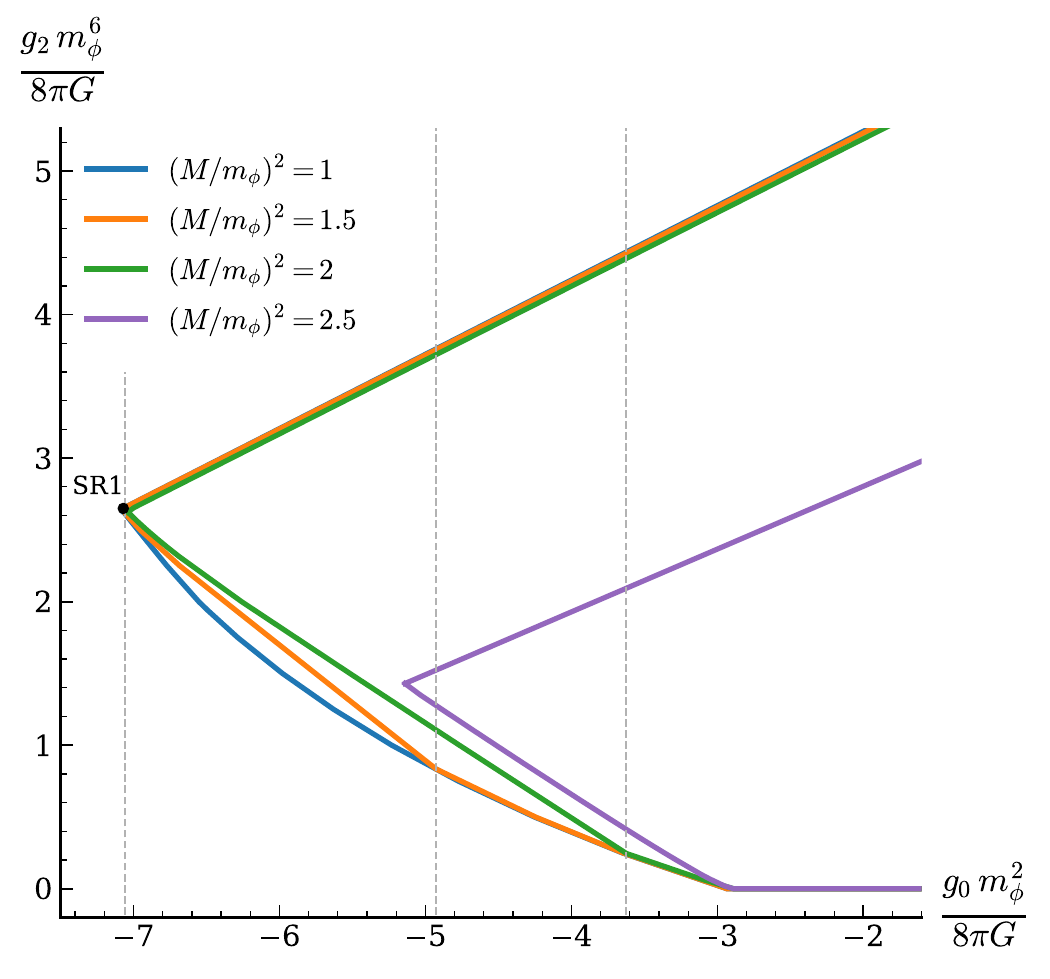}
    \caption{Exclusion plot in the $(\tilde g_0,\tilde g_2)$ plane at negative $\tilde g_0$ values for various cutoffs, this is a zoomed-in version of figure~\ref{fig:HGr_g0g2full}. For cutoffs $m_\phi^2\leq M^2\leq 2m_\phi^2$, the $\tilde g_0$ minimizing kink (black, SR1) and the connected linear upper bound on $\tilde g_2$ both remain. For higher cutoffs, the exclusion region shrinks quickly. As the cutoff is increased incrementally more of the curved lower bound on $\tilde g_2$ is removed. The dashed gray lines indicate the sampled $\tilde g_0$ values for subsequent figures \ref{fig:lowgspin2plots}, and \ref{fig:specP1-P3}.}
    \label{fig:Mp2low}
\end{figure}

\paragraph{Negative $\mathbf{\tilde g_0}$:}
We first focus on the negative $\tilde g_0$ region, shown in figure~\ref{fig:Mp2low} for several cutoffs.
As the cutoff increases from $M^2=m_\phi^2$ to $M^2=2m_\phi^2$, progressively larger portions of the curved boundary are excluded, while the $\tilde g_0$-minimizing point (SR1) remains unchanged. For $M^2>2m_\phi^2$, this point is also removed and the allowed region shrinks rapidly.

\begin{figure}[htbp]
    \centering
    \begin{subfigure}{0.49\textwidth}
        \centering
        \includegraphics[width=\linewidth]{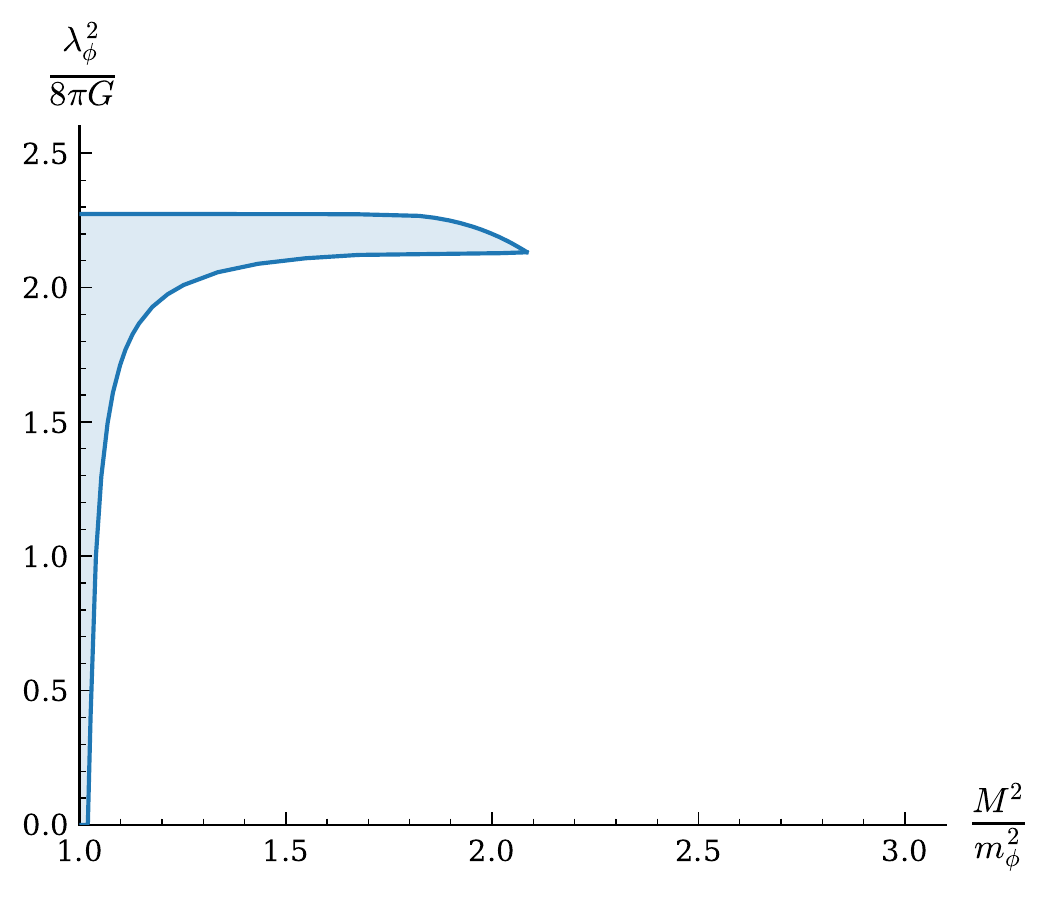}
        \caption{$\tilde g_0 = -7.06$}
        \label{fig:spin2P1}
    \end{subfigure}\\
    \begin{subfigure}{0.49\textwidth}
        \centering
        \includegraphics[width=\linewidth]{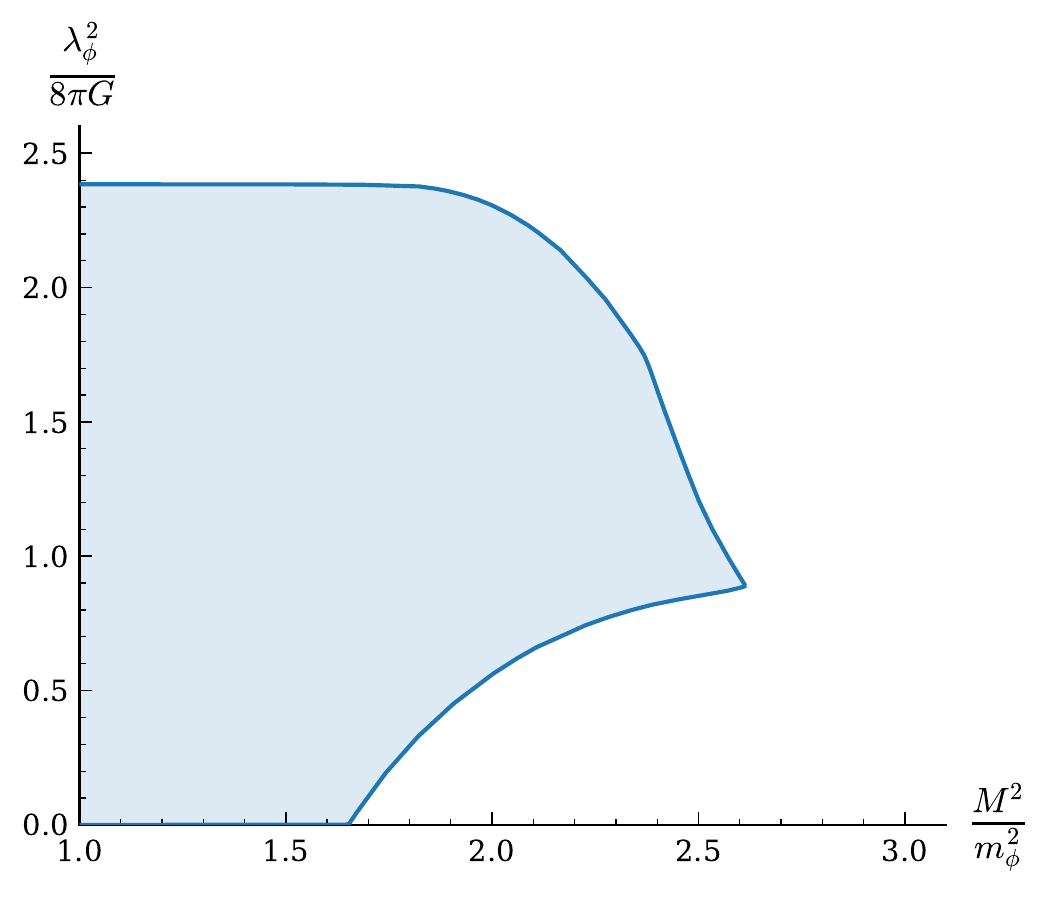}
        \caption{$\tilde g_0 = -4.9$}
        \label{fig:spin2P2}
    \end{subfigure}
    \hfill
    \begin{subfigure}{0.49\textwidth}
        \centering
        \includegraphics[width=\linewidth]{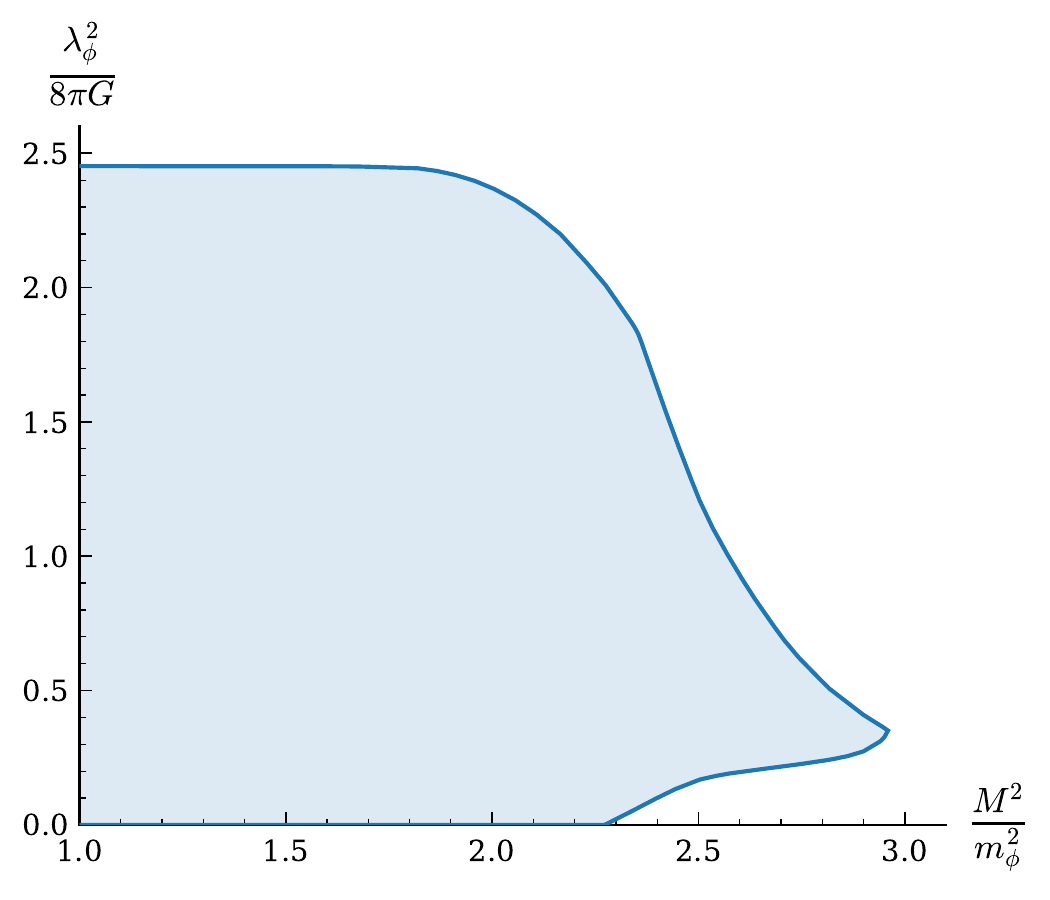}
        \caption{$\tilde g_0 = -3.6$}
        \label{fig:spin2P3}
    \end{subfigure}
    \caption{Upper and lower bounds on the coupling $\tilde \lambda_\phi^2$ of the spint-two exchange with mass $m_\phi$, shown as a function of the cutoff for three representative values of $\tilde g_0$. These values correspond to the gray dashed lines in figure~\ref{fig:Mp2low}. The blue shaded region denotes the allowed region. All three plots were generated with $k_{\text{max}}= 5$ and $n_{\text{max}}=11$.}
    \label{fig:lowgspin2plots}
\end{figure}

To probe the structure further, we fix $\tilde g_0$ at three representative values and bound the spint-two coupling at mass $m_\phi$, as shown in figure~\ref{fig:lowgspin2plots}. The chosen values correspond to the dashed gray lines in figure~\ref{fig:Mp2low}. A notable new feature is that one can obtain not only upper but also lower bounds on the spint-two coupling. For $\tilde g_0 <2$, amplitudes supported only by infinitely heavy states are excluded, which allows for a nontrivial lower bound on $\tilde \lambda_\phi$.

The structure of the bounds exhibits several notable features. For $\tilde g_0$ near its minimal allowed value (see figure~\ref{fig:spin2P1}), the allowed region is extremely narrow and persists only up to a cutoff $M^2 \approx 2 m_\phi^2$, indicating that the solution is highly constrained and effectively selects a unique amplitude. This solution must include an additional state at $m^2 = 2 m_\phi^2$.

For slightly higher values of $\tilde g_0$, shown in figures~\ref{fig:spin2P2} and \ref{fig:spin2P3}, the bounds exhibit a common qualitative structure. The upper bound remains approximately constant up to $M^2 \sim 2 m_\phi^2$, beyond which it decreases sharply. The lower bound is initially trivial, $\tilde \lambda_\phi^2 \geq 0$, up to a $\tilde g_0$-dependent cutoff, after which a kink develops and the lower bound increases until it intersects the upper bound at an intermediate point. As $\tilde g_0$ increases, this intersection occurs at smaller coupling and larger cutoff.

Comparing the bounds on the spint-two coupling in figures \ref{fig:spin2P2} and \ref{fig:spin2P3} with the corresponding slices of the exclusion region in figure~\ref{fig:Mp2low}, we can match features between the two descriptions. The upper bound on the spint-two coupling grows linearly with $\tilde g_0$, mirroring the linear upper boundary in $\tilde g_2$, and drops sharply at $M^2 = 2 m_\phi^2$ for all $\tilde g_0$ in this range. The lower bound becomes nontrivial at a $\tilde g_0$-dependent cutoff, which coincides with the point where the curved boundary in the $(\tilde g_0,\tilde g_2)$ plane develops a kink.

This kink has a clear physical interpretation. Its onset indicates that the extremal amplitude does not contain a spint-two exchange at $m_\phi$, and hence no resonances at all at this mass. Instead, the first states must appear at the cutoff scale where the kink develops.

Finally, the maximal allowed cutoff for fixed $\tilde g_0$ corresponds to the point where the solution moves into the interior of the exclusion region, approaching the configuration that minimizes $\tilde g_0$ at that cutoff.

One remark is in order regarding the numerics. The plots in figure~\ref{fig:lowgspin2plots} were generated with $k_{\text{max}} = 5$ and $n_{\text{max}} = 11$, whereas figure~\ref{fig:Mp2low} uses $k_{\text{max}} = 7$ and $n_{\text{max}} = 15$. This choice is motivated by the observation of \cite{Albert:2024yap} that, in the presence of spectral assumptions and for larger cutoffs, the bounds become increasingly sensitive to the large spin regime. Ensuring positivity in this regime requires greater care. To remain conservative while keeping the computation tractable, we therefore use a smaller number of wavepackets for the spint-two coupling bounds. As a result, the qualitative features are robust, although the precise numerical values may still shift. We expect improved quantitative agreement at higher truncation orders.

\begin{figure}[htbp]
    \centering
    \begin{subfigure}{0.32\textwidth}
        \centering
        \includegraphics[width=\linewidth]{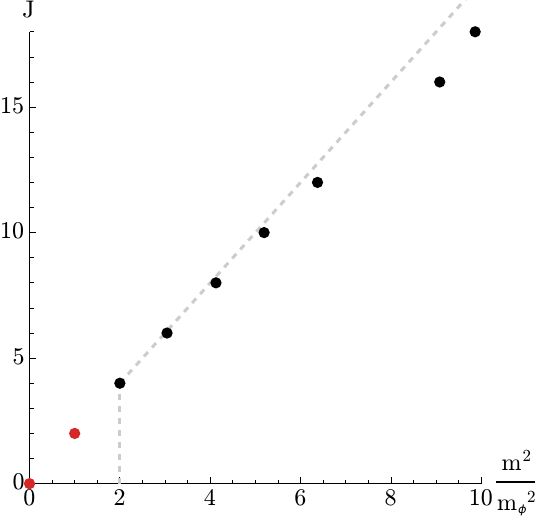}
        \caption{$\tilde g_0 = -7.06$}
        \label{fig:lowspinspec_P1}
    \end{subfigure}
    \hfill
    \begin{subfigure}{0.32\textwidth}
        \centering
        \includegraphics[width=\linewidth]{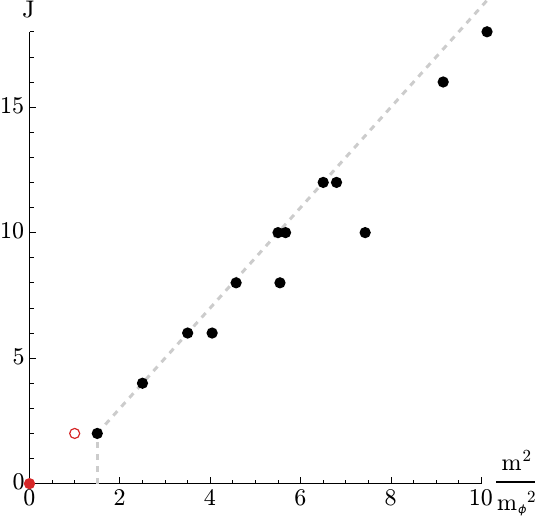}
        \caption{$\tilde g_0 = -4.9$}
        \label{fig:lowspinspec_P2}
    \end{subfigure}
    \hfill
    \begin{subfigure}{0.32\textwidth}
        \centering
        \includegraphics[width=\linewidth]{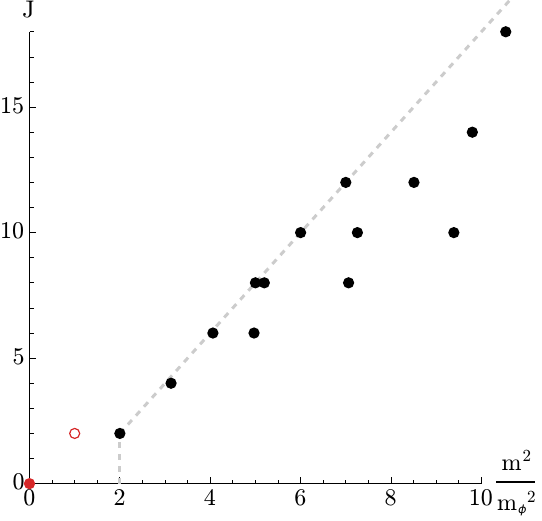}
        \caption{$\tilde g_0 = -3.6$}
        \label{fig:lowspinspec_P3}
    \end{subfigure}
    \caption{Chew--Frautschi plots for three representative spectra: figure \subref{fig:lowspinspec_P1} shows the globally $g_0$-minimizing point, while figures \subref{fig:lowspinspec_P2} and \subref{fig:lowspinspec_P3} correspond to points that minimize $g_2$ at fixed $g_0$. The chosen values of $g_0$ match the gray dashed lines in figure~\ref{fig:Mp2low}. The spectra are extracted from \texttt{SDPB} with additional spin-dependent lower bounds on the masses (indicated by gray dashed lines), chosen so as not to modify the bounds. Red markers denote the graviton pole at $(m=0,J=0)$ and the spint-two state at $m=m_\phi$, which sets the mass scale (shown as a ring for reference when the corresponding coupling vanishes). The spectra are extracted from \texttt{SDPB} and may still contain spurious points to the right of the leading trajectory.}
    \label{fig:specP1-P3}
\end{figure}

We can further analyze the structure of the extremal amplitudes along the curved portion of the lower bound on $\tilde g_2$, we extract the corresponding spectra from \texttt{SDPB}. As was discussed in the non-gravitational analysis, the raw spectra typically contain spurious states, which can be removed by imposing additional spin-dependent spectral constraints that leave the bound unchanged. In practice, we impose the maximal lower bounds on the masses (as a function of spin) that do not modify the solution. No assumptions are made on the spectrum from above. The resulting Chew–Frautschi plots for the same three representative values of $\tilde g_0$ are shown in figure~\ref{fig:specP1-P3}. 

At the $\tilde g_0$-minimizing point (figure~\ref{fig:lowspinspec_P1}), the spectrum is particularly simple and is well described by a single linear Regge trajectory passing through both the graviton pole and the first spint-two resonance. Notable, this trajectory coincides with the leading Regge trajectory of string theory.

For the other two cases, shown in figures~\ref{fig:lowspinspec_P2} and \ref{fig:lowspinspec_P3}, the spectra again exhibit a linear Regge trajectory always with the same slope. However, in these cases the trajectory is shifted so that the first state appears at spin $J=2$, with its mass set by the cutoff scale at which the kink develops in figure~\ref{fig:Mp2low}. In addition, both spectra contain heavier states that exhibit some structure, potentially indicative of subleading trajectories. However, states lying to the right of the leading trajectory have not been independently verified to correspond to physical excitations and may instead reflect residual artifacts of the numerical procedure. For the purposes of the present analysis, we therefore treat these higher-mass states as spurious and focus on the leading Regge trajectory in order to classify the family of extremal solutions, leaving a more detailed investigation of this structure for future work.

Taken together, these observations suggest that the curved boundary is generated by a continuous family of single Regge trajectories with fixed slope, where the mass of the first spint-two state increases as one moves toward $\tilde g_2 = 0$. In this limit, all resonances are pushed to infinite mass, consistent with the behavior inferred from the sum rules.

\begin{figure}[ht]
    \centering
    \includegraphics[width=0.85\textwidth]{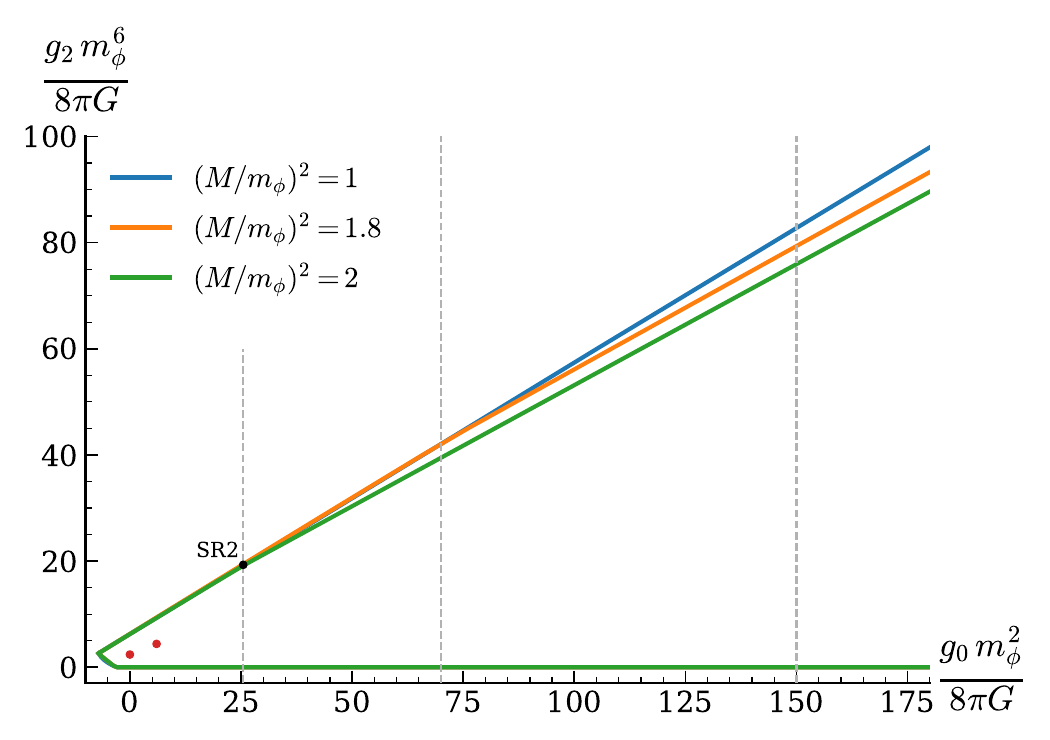}
    \caption{Exclusion plot in the $(\tilde g_0,\tilde g_2)$ plane, shown over an extended range of $\tilde g_0$ to highlight the large-$\tilde g_0$ region. This is a zoomed-out version of figure~\ref{fig:HGr_g0g2full} with three different values of the cutoff $M$. The red dots indicate the corresponding string theory values, included to provide a reference for the overall scale. For cutoffs satisfying $M^2>m_\phi^2$, the boundary exhibits a kink at a cutoff-dependent point where the slope decreases from the universal value $1/2$ to a smaller value consistent with the upper bound on $g_2 m_\phi^4/g_0$ obtained in the supersymmetric scalar analysis (for instance, $g_2 m_\phi^4/g_0 \approx 0.44$ for $M^2=2m_\phi^2$, marked by the black dot labeled SR2). The associated kink shifts toward $\tilde g_0 \to \infty$ as $M \to m_\phi$. The dashed gray lines indicate the sampled $\tilde g_0$ values for subsequent figures \ref{fig:P4-P6}, \ref{fig:specP4-P6}, and \ref{fig:specP456}.}
    \label{fig:Mp2high}
\end{figure}

\paragraph{Positive $\mathbf{\tilde g_0}$:}
We now turn to the regime of positive $\tilde g_0$. The exclusion regions for several values of the cutoff are shown in figure~\ref{fig:Mp2high}.
For cutoffs $m_\phi^2 < M^2 \leq 2m_\phi^2$, part of the linear boundary from the analysis without spectral assumptions persists. In this regime, the upper boundary initially has slope $1/2$, but develops a kink at a cutoff-dependent value of $\tilde g_0$. Beyond this point, the slope decreases to the corresponding value associated to the $g_2 m_\phi^4/g_0$ maximization in the non-gravitational system. For example, for the slope $M^2=2m_\phi^2$ the slope of the ray going out from the corresponding kink (SR2) matches the value ($\approx 0.44$) associated with the $g_3 M^6/g_0$-maximizing amplitude (see figure~\ref{fig:g2g3wspec}). As the cutoff is lowered, this kink moves to larger values of $\tilde g_0$, restoring the slope of 1/2 and the corresponding infinite spin tower at $M=m_\phi$. For $M^2 \geq 2m_\phi^2$, the boundary fully departs from the slope-$1/2$ line as we saw already in the negative $\tilde g_0$ plot (figure \ref{fig:Mp2low}). This behavior indicates that the non-gravitational infinite spin tower amplitude masks a second family of extremal gravitational solutions, which becomes visible only once the IST is excluded by the spectral assumption.

\begin{figure}[ht]
    \centering
    \includegraphics[width=0.7\textwidth]{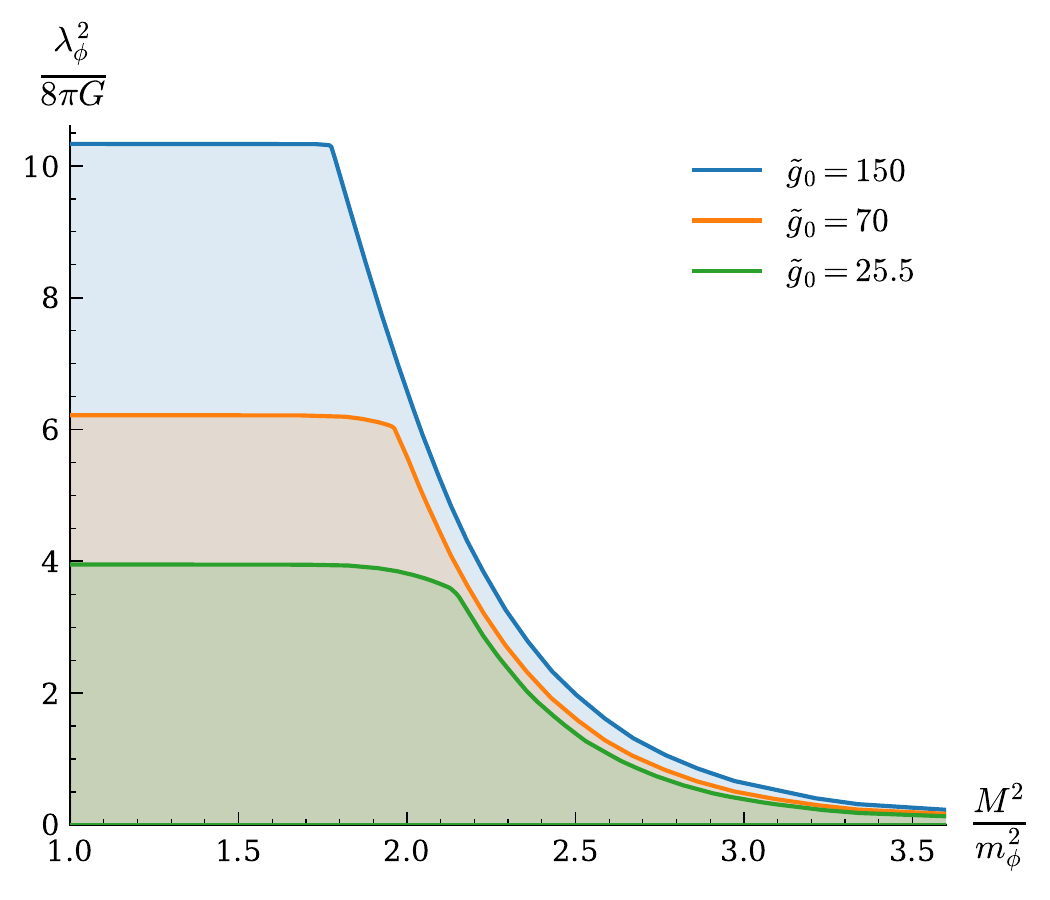}
    \caption{Upper bounds on the coupling $\tilde \lambda_\phi^2$ of the spint-two exchange with mass $m_\phi$, shown as a function of the cutoff for three representative values of $\tilde g_0$. These values correspond to the gray dashed lines in figure~\ref{fig:Mp2high}. In all cases, the lower bound is trivial, $\tilde \lambda_\phi^2 \geq 0$. The upper bounds exhibit a common structure: a maximal value that remains constant up to a cutoff scale set by $\tilde g_0$, beyond which the bound drops sharply. As $\tilde g_0$ increases, this kink shifts toward lower cutoff values. All three bounds were generated with $k_{\text{max}} = 5$ and $n_{\text{max}}=11$.}
    \label{fig:P4-P6}
\end{figure}

To further probe this structure, we fix $\tilde g_0$ and bound the spint-two coupling at mass $m_\phi$, as shown in figure~\ref{fig:P4-P6}. The chosen values of $\tilde g_0$ correspond to the dashed lines in figure~\ref{fig:Mp2high}.
In contrast to the negative $\tilde g_0$ case, the lower bound on the spint-two coupling is always trivial. This is because, for fixed positive $\tilde g_0$, a point with $\tilde g_2=0$ is always allowed and corresponds to amplitudes with no finite-mass resonances and hence $\lambda_\phi=0$. 

The upper bound exhibits a universal structure across all three values of $\tilde g_0$. The coupling is bounded by a maximal value that remains approximately constant up to a critical cutoff, beyond which it decreases rapidly and approaches zero as $M \to \infty$. As $\tilde g_0$ increases, this critical cutoff shifts to lower values. The three curves are nested, with larger values of $\tilde g_0$ enclosing those at smaller $\tilde g_0$, and the maximal coupling increasing approximately linearly with $\tilde g_0$. Despite this, the large-cutoff tails of the bounds nearly coincide (up to small numerical differences). This behavior is explained by the fact that one can increase $\tilde g_0$ by adding a decoupled non-gravitational sector, which shifts $\tilde g_0$ without modifying the coupling of the spint-two state at the first resonance. This behavior is explained by the fact that one can increase $\tilde g_0$ by adding a decoupled non-gravitational sector, which shifts $\tilde g_0$ without modifying the coupling of the spint-two state at the first resonance. The near overlap of the curves in this regime further indicates that no qualitatively new structure emerges in the bulk of the bounds as $\tilde g_0$ is increased.

\begin{figure}
    \centering
    \begin{subfigure}{0.32\textwidth}
        \centering
        \includegraphics[width=\linewidth]{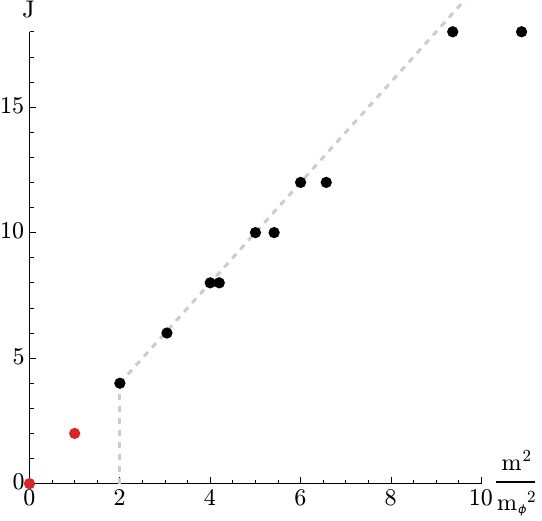}
        \caption{$\tilde g_0 = 25.5$}
    \end{subfigure}
    \hfill
    \begin{subfigure}{0.32\textwidth}
        \centering
        \includegraphics[width=\linewidth]{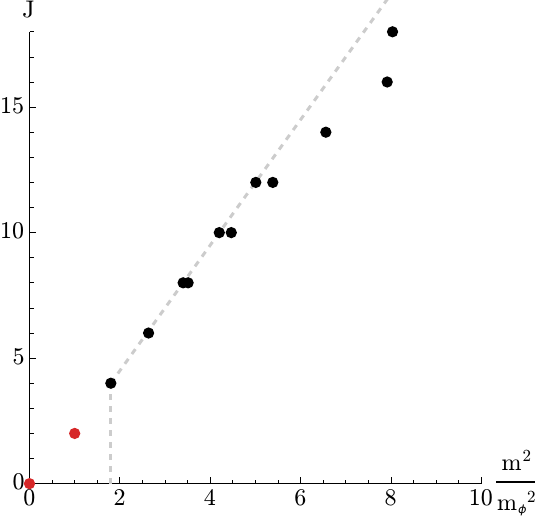}
        \caption{$\tilde g_0 = 70$}
    \end{subfigure}
    \hfill
    \begin{subfigure}{0.32\textwidth}
        \centering
        \includegraphics[width=\linewidth]{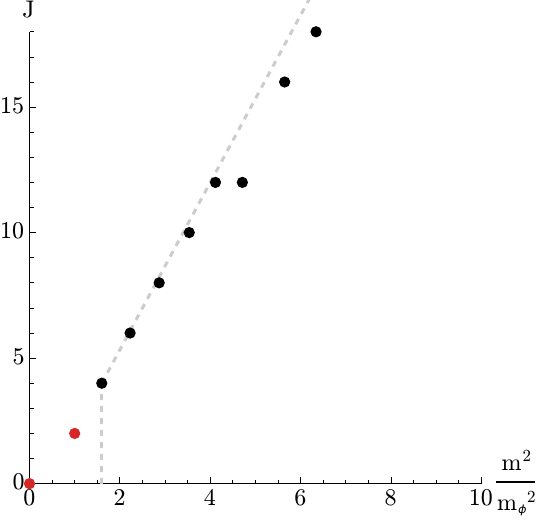}
        \caption{$\tilde g_0 = 150$}
    \end{subfigure}
    \caption{Chew--Frautschi plots for three representative spectra corresponding to points that maximize $\tilde g_2$ at fixed $\tilde g_0$. The chosen values of $\tilde g_0$ match the gray dashed lines in figure~\ref{fig:Mp2high}. The spectra are obtained under spin-dependent spectral assumptions, indicated by the gray dashed lines in the plots. The red markers denote the graviton exchange at $(m=J=0)$ and the spint-two exchange at $m=m_\phi$, which sets the mass scale. As the coupling increases, the slope of the leading Regge trajectory increases. The spectra are extracted from \texttt{SDPB} and may still contain spurious points to the right of the leading trajectory.}
    \label{fig:specP4-P6}
\end{figure}

\begin{figure}[ht]
    \centering
    \includegraphics[width=0.5\textwidth]{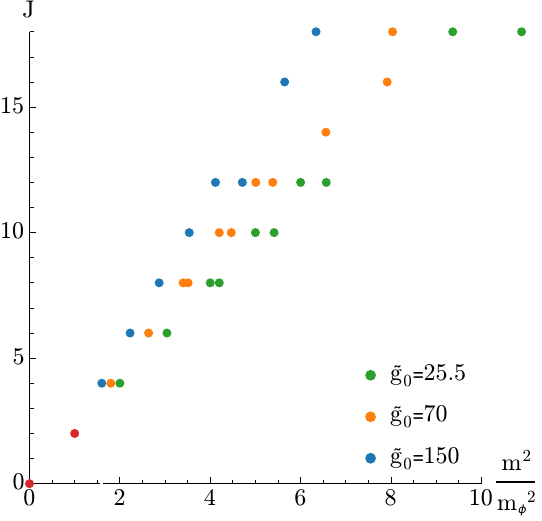}
    \caption{Overlay of the three Chew-Frautschi plots from figure~\ref{fig:specP4-P6}. As $\tilde g_0$ is increased the slope of the leading trajectory through the spint-two resonance at $m=m_\phi$ increases. As $\tilde g_0 \rightarrow \infty$, the spectrum will become an infinite spin tower decoupled from gravity, see \eqref{eq:towerofstates}.}
    \label{fig:specP456}
\end{figure}

We can again extract the extremal spectra from \texttt{SDPB} at the upper boundary, shown in figure~\ref{fig:specP4-P6}, together with an overlay for comparison in figure~\ref{fig:specP456}.
The extracted spectra are significantly cleaner in this regime, exhibiting well-defined single Regge trajectories that admit a straightforward interpretation and indicating another family of amplitudes. In this case, the first resonance remains at fixed mass $m_\phi$, while the slope increases with increasing $\tilde g_0$. This family interpolates between the tower-of-states limit at $\tilde g_0 \to \infty$ with $M = m_\phi$, and a second single-Regge-trajectory amplitude with slope and intercept matching that of string theory, albeit without subleading trajectories. Despite sharing the same spectrum, this latter amplitude is distinct from the one minimizing $\tilde g_0$, as their EFT expansions differ.

\paragraph{Summary:} 

The boundary of the exclusion region is generated by two distinct families of amplitudes, both whose spectrum forms a single Regge trajectory. The first family controls the boundary near the minimum value of $\tilde g_0$, while the second governs the large-$\tilde g_0$. Both families terminate in amplitudes whose leading Regge trajectories have the same slope and intercept as those appearing in string theory. The endpoints of these families, labeled SR1 and SR2 in the exclusion plots, are amplitudes with the same slope and intercept as the leading Regge trajectory in string theory amplitudes. 

It is noteworthy that similar families of extremal amplitudes have appeared in the maximally supersymmetric analyses of Type II and Type I strings \cite{Albert:2024yap}. In the setup realized by closed strings, the exclusion plot is spanned by single-Regge-trajectory families with fixed slope and varying intercept (along with the infinite spin tower amplitude \eqref{eq:towerofstates}). In contrast, in the open-string setup the allowed region is generated by families with fixed first resonance and varying slope, interpolating to the infinite spin tower configuration.  

These extremal solutions capture the structure of the boundary of the allowed space. One can move into the bulk of the allowed region, where the string theory amplitudes reside, by adding non-gravitational amplitudes. These act as subleading Regge trajectories, extending the spectrum beyond the extremal solutions. While they do not reproduce the full string spectrum, they bring the construction closer to the stringy structure and help explain why string theory sits in the interior of the exclusion region.

\section{Gluon scattering}\label{sec:gluons}

\subsection{Preliminaries}
We now turn our attention to the case of $2\to2$ scattering with external states belonging to the $D =~10$, $\mathcal{N} = (1,0)$ vector multiplet. This is the multiplet containing the gluons and its superpartners, which are valued in the adjoint of $G = SO(N)$. While  $N = 32$ is required for Green-Schwarz anomaly cancellation, we will keep expressions general, since this requirement is not necessary from the point of view of the bootstrap. When presenting concrete numerical results, we will however take $N = 32$ to allow for comparison with the heterotic string theory values.

\subsubsection{Super Ward identities}

The super Ward identities relevant to this case are well known and appear in various places in the literature, see for example \cite{Elvang:2013cua, Albert:2024yap}. One key difference is that we do not assume a planar limit as in \cite{Albert:2024yap,Berman:2023jys,Berman:2024wyt,Elvang:2026pmc}, so trace structures beyond the single-trace sector are allowed. For completeness, and to contrast with the gravitational case above, we briefly review the Ward identities. We work in the same on-shell superspace as before. Upon reduction to $D = 4$, the $D=10$, $\mathcal{N} = (1,0)$ vector multiplet becomes a single $\mathcal{N}=4$ multiplet, and the super Ward identities therefore admit a unique solution of the same form as in the previous case:
\begin{equation}\label{eq:superward}
    \mathcal{A}_{4V}^{abcd}(p_i,\eta_i) = \delta^{10}(p_1+p_2+p_3+p_4)\,\delta^8(\bar{Q})\,\mathcal{T}^{abcd}(s,u)\, .
\end{equation}
There is, however, one important distinction between the gauge and gravity cases. In the gravity case, the vector multiplets under consideration were Abelian, so there was no distinction between the external states, leading to maximal crossing symmetry of the amplitude. In the present case, the vector multiplets are non-Abelian, and the amplitudes therefore carry color indices $a,b,c,d$. Accordingly, the amplitude can be decomposed into the allowed trace structures of the gauge group generators. Explicitly, we write 
\small
\begin{equation}\label{eq:traceM}
    \begin{alignedat}{1}  
        \mathcal{T}^{abcd} = &\text{Tr}\!\left(\mbox{$ T^aT^bT^cT^d$}\right)\mathcal{M}^{(1)}(s,t) + \text{Tr}\!\left(\mbox{$T^aT^cT^dT^b$}\right)\mathcal{M}^{(1)}(s,u) + \text{Tr}\!\left(\mbox{$T^aT^dT^bT^c$}\right)\mathcal{M}^{(1)}(t,u) 
        \\
        & + \text{Tr}\!\left(\mbox{$T^aT^c$}\right)\!\text{Tr}\!\left(\mbox{$T^bT^d$}\right)\mathcal{M}^{(2)}(s,t) + \text{Tr}\!\left(\mbox{$T^aT^d$}\right)\!\text{Tr}\!\left(\mbox{$T^bT^c$}\right)\mathcal{M}^{(2)}(s,u) \\&+ \text{Tr}\!\left(\mbox{$T^aT^b$}\right)\!\text{Tr}\!\left(\mbox{$T^cT^d$}\right)\mathcal{M}^{(2)}(t,u)\, ,
    \end{alignedat}
\end{equation}
\normalsize
where $T^a$ are generators of $SO(N)$ in the fundamental representation\footnote{We follow the slightly unconventional normalization of \cite{Cvitanovic:2008zz}, with $\text{Tr}\,(T^aT^b)=\delta^{ab}$.}, with \newline $a = 1, \ldots, \frac{1}{2}N(N-1)$, and $\mathcal{M}^{(i)}$ are color-stripped amplitudes parameterizing the single- and double-trace contributions. The single-trace terms receive contributions from the gluon exchange, while the double-trace terms contain the graviton exchange. As before, $\eqref{eq:superward}$ contains amplitudes for scattering of all components of the vector superfield, and we may project onto specific external states. For the scattering of the helicity $\pm1$ components of the gauge field, $\phi_\pm$, we obtain
\begin{equation}\label{eq:GluonWardids}
    A^{abcd}(\phi_-\phi_- \to \phi_+\phi_+) = s^2\, \mathcal{T}^{abcd}(s,u)\, ,
\end{equation}
so that the Regge behavior of $\mathcal{M}^{(i)}$ is improved by two units.

\subsubsection{Unitarity: SO($N$) projection}\label{sec:projection}

Having fixed the supersymmetric kinematics, we now turn to imposing unitarity in the form of positivity constraints on the partial wave expansion of the amplitude. To this end, we decompose the amplitude into contributions from irreducible representations appearing in the $s$-channel of the tensor product of two SO($N$) adjoints:
\begin{equation}\label{eq:irrepM}
    \mathcal{T}^{abcd} = \sum_R \mathcal{M}^R(s|t,u)\, [\mathbb{P}_R]^{abcd} \,,
\end{equation}
where the sum runs over six representations \cite{Cvitanovic:2008zz}, with projectors
\begin{align}
    \begin{split}
        [\mathbb{P}_1]^{abcd} &=  \frac{2}{N(N-1)} \delta^{ab}\delta^{cd}\, ,
        \\
        [\mathbb{P}_2]^{abcd} &=  \frac{4}{N-2}\left(d^{abcd}+\frac{1}{12}\left(f^{adi}f^{bci}-f^{aci}f^{dbi}\right)-\frac{1}{N}\delta^{ab}\delta^{cd}\right)\, ,
        \\
        [\mathbb{P}_3]^{abcd} &=  \frac{2}{3}d^{abcd} + \frac{1}{9}\left(f^{adi}f^{bci}+f^{aci}f^{dbi}\right) + \frac{1}{3}\left(\delta^{ad}\delta^{bc}+\delta^{ac}\delta^{bd}\right)-[\mathbb{P}_1]^{abcd}-[\mathbb{P}_2]^{abcd}\, ,
        \\
        [\mathbb{P}_4]^{abcd} &=  -\frac{2}{3}d^{abcd} + \frac{1}{9}\left(f^{adi}f^{bci}-f^{aci}f^{dbi}\right) + \frac{1}{6}\left(\delta^{ad}\delta^{bc}+\delta^{ac}\delta^{bd}\right)\, ,
        \\
        [\mathbb{P}_5]^{abcd} &=  \frac{1}{N-2}f^{abi}f^{dci}\, ,
        \\
        [\mathbb{P}_6]^{abcd} &=  \frac{1}{2}\left(\delta^{ad}\delta^{bc}-\delta^{ac}\delta^{bd}\right)-[\mathbb{P}_5]^{abcd}\, ,
    \end{split}
\end{align}
satisfying
\begin{equation}
    [\mathbb{P}_R]^{abef} \, [\mathbb{P}_{R'}]^{efcd}=\delta_{R,R'}\, [\mathbb{P}_R]^{abcd} \quad \text{and} \quad \sum_{R=1}^{6}[\mathbb{P}_R]^{abcd} = \delta^{ad}\delta^{bc}.
\end{equation}
In this numbering $R=1$ denotes the singlet and $R=5$ the adjoint representation. Note the representations 1-4 are symmetric while 5 and 6 are anti-symmetric. Here $f^{abc}$ are the usual structure constants and $d^{abcd}$ denotes the fully antisymmetric tensor defined by
\begin{equation}
    \text{Tr}\,(T^aT^bT^cT^d) = d^{abcd} + \frac{1}{6}\left(f^{adi}f^{bci} - f^{abi}f^{cdi}\right).
\end{equation}
The two parameterizations, \eqref{eq:traceM} and \eqref{eq:irrepM}, are related as follows (separating the single- and double-trace contributions for convenience, i.e. $\mathcal{M}^R=\mathcal{M}^{R,(1)}+\mathcal{M}^{R,(2)}$):
\begin{align}\label{eq:RepBasis}
    \begin{split}
        \mathcal{M}^{1,(1)}(s|t,u) &= \frac{N-1}{2}\big(\mathcal{M}^{(1)}(s,t)+\mathcal{M}^{(1)}(s,u)\big) + \frac{1}{2}\mathcal{M}^{(1)}(t,u)\, ,
        \\
        \mathcal{M}^{2,(1)}(s|t,u) &= \frac{N-2}{4}\big(\mathcal{M}^{(1)}(s,t)+\mathcal{M}^{(1)}(s,u)\big) + \frac{1}{2}\mathcal{M}^{(1)}(t,u)\, ,
        \\
        \mathcal{M}^{3,(1)}(s|t,u) &= \frac{1}{2}\mathcal{M}^{(1)}(t,u)\, ,
        \\
        \mathcal{M}^{4,(1)}(s|t,u) &= -\mathcal{M}^{(1)}(t,u)\, ,
        \\
        \mathcal{M}^{5,(1)}(s|t,u) &= \frac{N-2}{4}\big(\mathcal{M}^{(1)}(s,u)-\mathcal{M}^{(1)}(s,t)\big)\, ,
        \\
        \mathcal{M}^{6,(1)}(s|t,u) &= 0\, ,
    \end{split}
\end{align}
and
\begin{align}\label{eq:RepBasis2}
    \begin{split}
        \mathcal{M}^{1,(2)}(s|t,u) &= \frac{N(N-1)}{2}\mathcal{M}^{(2)}(t,u) + \mathcal{M}^{(2)}(s,t) + \mathcal{M}^{(2)}(s,u)\, ,
        \\
        \mathcal{M}^{2,(2)}(s|t,u) &= \mathcal{M}^{(2)}(s,t) + \mathcal{M}^{(2)}(s,u)\, ,
        \\
        \mathcal{M}^{3,(2)}(s|t,u) &= \mathcal{M}^{(2)}(s,t) + \mathcal{M}^{(2)}(s,u)\, ,
        \\
        \mathcal{M}^{4,(2)}(s|t,u) &= \mathcal{M}^{(2)}(s,t) + \mathcal{M}^{(2)}(s,u)\, ,
        \\
        \mathcal{M}^{5,(2)}(s|t,u) &= \mathcal{M}^{(2)}(s,t) - \mathcal{M}^{(2)}(s,u)\, ,
        \\
        \mathcal{M}^{6,(2)}(s|t,u) &= \mathcal{M}^{(2)}(s,t) - \mathcal{M}^{(2)}(s,u)\, ,
    \end{split}
\end{align}
where $R=1,\dots,6$ labels the representation.

Unitarity requires positivity of the spectral densities in each representation channel, i.e. 
\begin{equation}\label{eq:rhoRpositivity}
    \rho_J^R(s)\geq 0 \quad  \text{in the physical region} \quad (s>0\, ,u<0)
\end{equation} 
appearing in the spectral decomposition of $\mathcal{M}^R$:
\begin{equation}\label{eq:PWER}
    {\rm Im}[\mathcal{M}^R(s,u)] = s^{\frac{4-D}{2}} \sum_{J} n^{(D)}_J \rho^R_J(s) \, \mathcal{P}_J\!\left(\mbox{$1+\frac{2u}{s}$}\right),
\end{equation}
where the sum over spins is restricted to even spins for the symmetric representations (1-4) and odd spins for the antisymmetric representations (5-6).

\subsubsection{Other assumptions}

In addition to supersymmetry and unitarity, we will impose crossing symmetry, Regge boundedness, as well as a well-behaved low energy EFT limit.
The crossing properties of the $\mathcal{M}^{(i)}$ follow directly from the decomposition \eqref{eq:traceM} and the anti-symmetry of the generators, 
\begin{equation}
    \text{Tr}\,\left(\mbox{$T^aT^bT^cT^d$}\right) = \text{Tr}\,\left(\mbox{$T^aT^dT^cT^b$}\right),
\end{equation} 
implying both the single and double trace components of the amplitude enjoy ($s\leftrightarrow u$) crossing symmetry:
\begin{equation}\label{eq:M1&2sym}
    \mathcal{M}^{(1)}(s,u) = \mathcal{M}^{(1)}(u,s) \quad \text{and} \quad \mathcal{M}^{(2)}(s,u) = \mathcal{M}^{(2)}(u,s).
\end{equation}
Our assumption regarding Regge boundedness is that massless exchanges reggeize in the full theory. Since $\mathcal{M}^{(1)}$ and $\mathcal{M}^{(2)}$ contain, respectively, gluon and graviton exchange, we expect them to exhibit spin-1 and spint-two Regge behavior. The overall factor of $s^2$ implied by the super Ward identities \eqref{eq:GluonWardids} then reduces the effective Regge spin by two units, leading to asymptotic behavior corresponding to spin-$(-1)$ and spin-zero, respectively. Explicitly, we will assume
\begin{equation}\label{eq:gluonRegge}
    \lim_{|s|\to\infty}s\,\mathcal{M}^{(1)}(s,u) \to 0 \quad \text{and} \quad \lim_{|s|\to\infty}\mathcal{M}^{(2)}(s,u) \to 0 \, ,
\end{equation}
for fixed $u<0$. This assumption is stronger than the most conservative one, which would be to assume spint-two Regge behavior in both amplitudes. As we will see, this extra assumption yields a sum rule involving the gauge coupling $g^2_{YM}$ which cannot be used in the standard bootstrap setup. Our results are thus independent of this assumption and hold even if one were to assume the most conservative Regge behavior for the single trace amplitude. We include it nonetheless to later discuss the $g^2_{YM}$ sum rule.

So far, all consistency conditions have been imposed on the amplitudes $\mathcal{M}^{(i)}$ at arbitrary energies. As in the gravitational case, we now introduce additional assumptions that define the low-energy regime. For both amplitudes we assume a mass gap $M$ between the massless states and the first excited state(s), so that below this scale the dynamics are well described by an effective field theory of the massless degrees of freedom, namely the gluons, gravitons, and their superpartners. This implies that, in a finite neighborhood of the origin, the amplitudes admit the most general low-energy expansion consistent with $(s\leftrightarrow u)$ crossing symmetry,
\begin{align}
    \begin{split}
        \label{eq:MIRvec}
        \mathcal{M}^{(1)}_{\text{IR}}(s,u) &= -\frac{g^2_{\text{YM}}}{su} + g^{(1)}_0 + g^{(1)}_1(s+u) + \cdots\, , 
        \\
        \mathcal{M}^{(2)}_{\text{IR}}(s,u) &= -\frac{8\pi G}{t} + g^{(2)}_0 + g^{(2)}_1(s+u) + \cdots\, .
    \end{split}
\end{align}
The pole in $\mathcal{M}^{(1)}_{\text{IR}}$ corresponds to gluon exchange and matches the familiar structure from the planar limit. For $\mathcal{M}^{(2)}_{\text{IR}}$, the graviton exchange appears as an effective scalar $t$-channel pole. This can be understood most directly by decomposing the amplitude into representation channels, see \eqref{eq:RepBasis} and \eqref{eq:RepBasis2}. A singlet exchange that does not appear in other channels must arise from a pole in $\mathcal{M}^{(2)}(t,u)$. A similar argument applies to the gluon exchange, which transforms in the adjoint representation and therefore must contribute equally but with opposite sign to $\mathcal{M}^{(1)}(s,u)$ and $\mathcal{M}^{(1)}(s,t)$. 

As in the gravitational case, these massless poles are marginal with respect to the assumed Regge boundedness. The remaining terms in \eqref{eq:MIRvec} correspond to derivative corrections, parametrized by the Wilson coefficients $g^{(i)}_n$, and constitute the most general expansion consistent with crossing symmetry.

\subsubsection{String theory amplitudes} 

There are two known UV-complete amplitudes that satisfy all the assumptions outlined above. They arise from gluon scattering in heterotic and Type I string theory. In both cases the amplitudes realize the full set of consistency conditions, including crossing symmetry, unitarity, and Regge behavior.

We begin with the heterotic string, which provides the most general realization. In this case, all six representation channels contribute, and both single- and double-trace sectors are present. The amplitudes take the form \cite{green1987superstring1}
\begin{equation}\label{eq:hetamplitudes}
        \mathcal{M}_{\text{Het}}^{(1)}(s,u)=2\alpha't\, \mathcal{M}_{\text{VS}}(s,u) \quad \text{and} \quad \mathcal{M}_{\text{Het}}^{(2)}(s,u)=\frac{\alpha'^2su}{8(1+\alpha't/4)}\,\mathcal{M}_{\text{VS}}(s,u)\, ,
\end{equation}
where $\mathcal{M}_\text{VS}$ is the Virasoro--Shapiro amplitude \eqref{eq:VSamplitude}. Both amplitudes admit low energy expansions of the form \eqref{eq:MIRvec}, with coefficients
\begin{equation}
    \begin{alignedat}{3}
        g^2_{\text{YM}}=\frac{128}{\alpha'^2}&\,, \quad &&g_0^{(1)} = 0\,, \quad &&g_1^{(1)} =  4\zeta(3)\alpha' \, ,
        \\
        8\pi G = \frac{8}{\alpha'}&\,, \quad &&g_0^{(2)} = 2\,,\quad &&g_1^{(2)} = \frac{\alpha'}{2}\, .
    \end{alignedat}
\end{equation}
We now turn to the Type I $SO(32)$ string, which provides a particularly simple realization. In this case the amplitude is planar, so the double-trace sector vanishes identically. As a consequence, the representation-channel decomposition simplifies: requiring positivity of the spectral densities in the six representation channels \eqref{eq:RepBasis} implies that $\mathcal{M}^{(1)}(t,u)$ cannot contain any poles. Therefore, the only nontrivial contributions arise from the symmetric and antisymmetric combinations of $\mathcal{M}^{(1)}(s,u)$ and $\mathcal{M}^{(1)}(s,t)$. These correspond to even- and odd-spin exchanges, respectively, and can therefore be combined into a single spectral density with support over all integer spins. In this sense, the matrix-valued sum rules that will arise in the next section reduce to a single scalar system in the Type I case and were studied in \cite{Albert:2024yap}. The amplitude is the Veneziano amplitude:
\begin{equation}
    \mathcal{M}^{(1)}_{\text{Type I}}(s,u) = -\frac{\Gamma(-\frac{\alpha'}{2}s)\Gamma(-\frac{\alpha'}{2}u)}{\Gamma(1+\frac{\alpha'}{2}t)}\quad \text{and}\quad \mathcal{M}_{\text{Type I}}^{(2)}(s,u)=0\, .
\end{equation}
Matching to the low energy expansion yields
\begin{equation}
    g^2_{\text{YM}} = \frac{4}{\alpha'^2}\, , \quad g_0^{(1)} = \zeta(2)\, , \quad g_1^{(1)} = \frac{\alpha'}{2}\zeta(3)\, .
\end{equation} 
Finally, note that in the heterotic case the first massive resonance appears at $m^2 = 4/\alpha'$, whereas in the Type I normalization it appears at $m^2 = 2/\alpha'$. This difference is purely conventional and can be absorbed by a rescaling of $\alpha'$ if desired.

\subsection{Sum rules and null constraints}

The gluon bootstrap admits a more intricate set of dispersive sum rules than the gravitational case, due both to the presence of multiple color structures and the reduced crossing symmetry. The full amplitude decomposes into single-and double-trace components, $\mathcal{M}^{(1)}$ and $\mathcal{M}^{(2)}$ respectively, each with its own EFT expansion whose coefficients we would like to constrain. The two amplitudes are, however, not independent. Positivity is naturally formulated in the representation basis introduced in section \ref{sec:projection}, which mixes the two trace structures.

For each $\mathcal{M}^{(i)}$ we obtain two sets of sum rules, one at fixed $u$ and one at fixed $t$. Applying the standard contour argument in the complex $s$-plane therefore leads to four families of sum rules:
\begin{align}
    \begin{split}
        \mathcal{U}^{(i)}_k :& \qquad \frac{1}{2\pi i}\oint_\infty \frac{ds}{s}\frac{\mathcal{M}^{(i)}(s,u)}{s^k} = 0 
        \\
        \mathcal{V}_k^{(i)}:& \qquad \frac{1}{2\pi i}\oint_\infty\frac{ds}{s}\frac{\mathcal{M}^{(i)}(s, t)}{s^k} = 0\, .
    \end{split}
\end{align}
The contour at infinity vanishes provided sufficiently many subtractions are performed. The required number depends on the Regge behavior of the corresponding amplitude, which differs for $\mathcal{M}^{(1)}$ and $\mathcal{M}^{(2)}$ because of their different massless exchanges. From \eqref{eq:gluonRegge}, we find the allowed range is:
\begin{align}
    k =
        \begin{cases}
            -1,0,1,\ldots & \text{for } i = 1 
            \\
            0,1,2,\ldots & \text{for } i = 2\, .
        \end{cases}
\end{align}
Shrinking the contour relates the low energy EFT coefficients to high energy averages. After mapping the cuts to positive $s=m^2$ one sees that the fixed-$u$ sum rules capture $s$- and $t$-channel exchanges while the fixed $t$-sum rule picks up the $u$-channel poles.
\begin{align}
    \begin{split}
        \mathcal{U}^{(i)}_k : \quad \underset{s=0,-u}{\mathrm{Res}}\left[\frac{\mathcal{M}^{(i)}_{\text{IR}}(s,u)}{s^k}\right] &= \frac{1}{\pi}\int\displaylimits_{M^2}^\infty \frac{dm^2}{m^2}\left(\frac{\text{Im}[\mathcal{M}^{(i)}(s,u)]}{m^{2k}} + (-1)^k \frac{\text{Im}[\mathcal{M}^{(i)}(t,u)]}{(m^2+u)^{k+1}}\right)
        \\
        \mathcal{V}_k^{(i)}: \,\quad \underset{s=0,-u}{\mathrm{Res}}\left[\frac{\mathcal{M}^{(i)}_{\text{IR}}(s,t)}{s^k}\right]&= \frac{1}{\pi}\int\displaylimits_{M^2}^\infty \frac{dm^2}{m^2}\,\left(\frac{1}{m^{2k}} + \frac{(-1)^k m^2}{(m^2+u)^{k+1}}\right)\text{Im}[\mathcal{M}^{(i)}(s,t)]\, .
    \end{split}
\end{align}
Using the relations in \eqref{eq:RepBasis} and \eqref{eq:RepBasis2}, one can express $\mathcal{M}^{(i)}(s,u)$, $\mathcal{M}^{(i)}(t,u)$, and $\mathcal{M}^{(i)}(s,t)$ in terms of the six representation channels, whose spectral densities $\rho^R_J(s)$ are positive by unitarity, see \eqref{eq:rhoRpositivity}. We therefore introduce the positive semidefinite spectral density matrix
\begin{equation} 
    \bbrho_J(m^2) = \text{diag}\left(\rho^R_J(m^2)\right) \quad \text{with} \quad R=1,\ldots,6\, ,
\end{equation}
where $\text{diag}(\cdots)$ denotes a diagonal matrix. The corresponding matrix-valued heavy average is defined by
\begin{equation}\label{eq:gluonspectralavg}
    \left<(\cdots)\right>\equiv\frac{1}{\pi}\sum_{J}  n^{(D)}_J\int\displaylimits_{M^2}^\infty\frac{\mathrm{d}m^2}{m^2} m^{4-D}\,\text{Tr}\!\left(\bbrho_J(m^2)\,(\cdots)\right)\, .
\end{equation}
The allowed spins appearing in the sum depend on the representation channel, see \eqref{eq:PWER}. 

In terms of this heavy average, the lowest subtracted sum rule takes the form
\begin{equation}\label{eq:gYMsumrule}
    \mathcal{U}^{(1)}_{-1} : \quad -\frac{g^2_{YM}}{u} = \left<\frac{m^2}{15}\mathcal{P}_J\!\left(\mbox{$1+\frac{2u}{m^2}$}\right) \text{diag}(0\,,1\,,-11\,,10\,,1\,,-1)\right>\, ,
\end{equation}
which captures the gluon exchange in $\mathcal{M}^{(1)}$. There are no other sum rules at this order: $\mathcal{V}^{(1)}_{-1}$ vanishes identically, while the Regge behavior of $\mathcal{M}^{(2)}$ requires at least one further subtraction.

At the next order, four independent sum rules appear. The graviton exchange contributes to $\mathcal{V}_0^{(2)}$ and can be isolated by the combining it with the forward limit of higher sum rules, exactly as in the gravity sum rule \eqref{eq:SRgravG0}. This yields the improved sum rule
\begin{equation}\label{eq:gluonGsumrule}
    \mathcal{V}^{(2)}_{0,\text{ improved}}=\mathcal{V}^{(2)}_{0} - \sum_{k=0}^{\infty} (-u)^k\,\mathcal{U}^{(2)}_k(u = 0) : \qquad -\frac{8\pi G}{u} =\langle \mathbb{D}(u)\rangle\, ,
\end{equation}
where the matrix $\mathbb{D}(u)$ is diagonal, with entries
\begin{align}\label{eq:matrixD}
    \begin{split}
        \mathbb{D}_{11} &= -\frac{m^2}{496\,(m^2 - u)}\, , 
        \\[4pt]
        \mathbb{D}_{22} &= \frac{m^2}{240\,(m^2 - u)}\, , 
        \\[6pt]
        \mathbb{D}_{33} &=\frac{(2m^2+u)}{3(m^2+u)}\mathcal{P}_J\!\left(\mbox{$1 + \frac{2u}{m^2}$}\right)-\frac{2m^2(78m^2-77u)}{465(m^4-u^2)}\, , 
        \\[10pt]
        \mathbb{D}_{44} &=\frac{1}{6}\left(\frac{(2m^2+u)}{(m^2+u)}\mathcal{P}_J\!\left(\mbox{$1 + \frac{2u}{m^2}$}\right)-\frac{m^2}{(m^2+u)}\right)\, , 
        \\[10pt]
        \mathbb{D}_{55} &= 0\, , 
        \\[4pt]
        \mathbb{D}_{66} &=\frac{1}{2}\left(\frac{(2m^2+u)}{(m^2 + u)}\mathcal{P}_J\!\left(\mbox{$1 + \frac{2u}{m^2}$}\right) + \frac{m^2}{(m^2 + u)}\right).
    \end{split}
\end{align}
Unlike for graviton scattering, the improvement procedure isolates $G$, without mixing with any additional EFT coefficients.

The sum rules \eqref{eq:gYMsumrule} and \eqref{eq:gluonGsumrule} both contain poles at $u=0$ and so cannot be expanded in the forward limit. As in the gravitational case, they must instead be smeared against suitable wavepackets \cite{Caron-Huot:2021rmr,Albert:2024yap}, see section \ref{sec:gravity}. The remaining three sum rules at this order are regular at $u=0$ and admit a straightforward forward expansion:
\begin{align}\label{eq:forwardlevel0}
    \begin{split}
        \mathcal{U}^{(1)}_0(u=0)&: \qquad g^{(1)}_0 = \left<\frac{1}{15}\,\text{diag}(0\,,1\,,9\,,-10\,,1\,,-1) \right> 
        \\
        \mathcal{V}^{(1)}_0(u=0)&: \qquad g^{(1)}_0 = \left<\frac{1}{15}\,\text{diag}(0\,,2\,,-2\,,0\,,-2\,,2) \right> 
        \\
        \mathcal{U}_0^{(2)}(u=0)&: \qquad g^{(2)}_0 = \left<\frac{1}{7440}\, \text{diag}(15\,,-31\,,2496\,,1240\,,0\,,-3720)\right>
    \end{split}
\end{align}
As in \eqref{eq:gluonGsumrule}, one may alternatively construct improved versions of these sum rules by combining them with an infinite tower of null constraints. We record the corresponding expressions in App. \ref{app:extraSR}. The existence of two independent sum rules for $g_0^{(1)}$ implies a nontrivial null constraint. Their difference yields the lowest-order example, $\mathcal{Y}_{0,0}^{(1)}$, which we now define more generally.

In addition to the sum rules, there are four families of null constraints generated by the double contour integral \cite{Albert:2023jtd}:
\begin{equation}
    \oint_0\frac{du}{2\pi i}\oint_\infty\frac{ds}{2\pi i}\frac{1}{su} \left( \frac{\mathcal{M}^{(i)}(s,u\text{ or }t)}{s^{n-\ell}u^\ell} -\frac{\mathcal{M}^{(i)}(u\text{ or }t,s)}{u^{n-\ell}s^\ell}\right) = 0\, .
\end{equation}
These take the form
\begin{equation}
    \left< \mathcal{X}^{(i)}_{n,\ell}\right> = 0 \quad \text{and} \quad \left<\mathcal{Y}^{(i)}_{n,\ell} \right> = 0\, ,
\end{equation}
corresponding respectively to fixed $u$ and $t$ contour integrals, with
\begin{align}
    \begin{split}
        \mathcal{X}^{(i)}_{n,\ell} &= \mathop{\mathrm{Res}}_{u = 0}\left[\frac{1}{u}\left( \!\left( \frac{M^{(i)}(s,u)}{m^{2(n-\ell+1)}u^\ell}-\frac{M^{(i)}(s,u)}{u^{n-\ell}m^{2\ell+2}}\right)-\left(m^2\rightarrow -m^2-u,s\rightarrow t\right)\!\right)m^2\,\mathcal{P}_J\!\left(\mbox{$1+\tfrac{2u}{m^2}$}\right)\right] 
        \\
        \mathcal{Y}^{(i)}_{n,\ell} &= \mathop{\mathrm{Res}}_{u = 0}\left[\frac{1}{u}\left( \!\left( \frac{M^{(i)}(t,u)}{m^{2(n-\ell+1)}u^\ell}-\frac{M^{(i)}(s,t)}{u^{n-\ell}m^{2\ell+2}}\right)-\left(m^2\rightarrow -m^2-u, s\leftrightarrow t\right)\!\right)m^2\,\mathcal{P}_J\!\left(\mbox{$1+\tfrac{2u}{m^2}$}\right)\right]\, .
    \end{split}
\end{align}
As before, the trace-basis amplitudes must be rewritten in the representation basis using \eqref{eq:RepBasis} and \eqref{eq:RepBasis2}. The allowed ranges of $(n,\ell)$ depend on the family of constraints and are summarized in table 
\ref{tab:nandlvalues}. 
\begin{table}[ht]
    \makegapedcells
    \centering
    \begin{tabular}{c|c|c}
        Constraint family & minimum $n$ & $\ell$ range \\
        \hline
        $\mathcal{X}^{(i)}_{n,\ell}$ & $n\geq 1$ & $0\leq \ell \leq [ \frac{n-1}{2}]$ 
        \\
        $\mathcal{Y}^{(1)}_{n,\ell}$ & $n\geq 0$ & $0\leq \ell \leq [ \frac{n}{2}]$ 
        \\
        $\mathcal{Y}^{(2)}_{n,\ell}$ & $n\geq 2$ & $1\leq \ell \leq [ \frac{n}{2}]$
    \end{tabular}
    \caption{Ranges for independent null constraints.}
    \label{tab:nandlvalues}
    \end{table}
The shifted range for the $\mathcal{Y}^{(2)}_{n,\ell}$ family originates from the graviton pole in the $t$-channel of the double-trace amplitude, or equivalently in $\mathcal{V}_0^{(2)}$. Finally, we record the leading null constraint in powers of $m^2$:
\begin{equation}\label{eq:topNCs}
    \mathcal{Y}^{(1)}_{0,0} = \text{diag}(0\,,1\,, -11\,, 10\,, -3\,, 3)\, .
\end{equation}
As noted above, this constraint can equivalently be obtained by taking the difference of the two $g_0^{(1)}$ sum rules in \eqref{eq:forwardlevel0}.

\subsection{Obstructions to the usual procedure}

Having derived the full set of sum rules and null constraints, one would ordinarily proceed by defining a semidefinite programming problem and deriving bounds on ratios of EFT couplings. Schematically, one seeks bounds of the form
\begin{equation}
    A \leq \frac{g_{\text{obj}}\,M^\#}{g_{\text{norm}}} \leq B\, ,
\end{equation}
where $g_{\rm norm}$ is used to normalize the spectrum. For such bounds to exist, the normalization condition must be strictly positive\footnote{Strict positivity of the normalization condition is somewhat stronger than what was required in the open-string analysis of \cite{Albert:2024yap}. There, the normalization kernel was not positive for all masses and spins, but the negative region was confined to odd spins and low masses, and shrank rapidly as the spin increased.}. Otherwise, one can keep $g_{\rm norm}$ fixed while changing the quantity to be bounded, $g_{\text{obj}}$, arbitrarily by adding suitable spectral weight, rendering the optimization region non-compact and the resulting bounds trivial.

In the present system, the obstruction is immediate. The gluon- and graviton exchange sum rules are matrix-valued, and the corresponding matrices contain entries of mixed sign across the six representation channels. As a result, the associated high-energy averages are not positive definite. One can increase the quantity one wishes to bound arbitrarily by adding spectral weight in one channel, while compensating the normalization by adding weight in another channel with the opposite sign. This is closely analogous to the standard obstruction preventing one from bounding quantities at lower mass order than the normalization condition: contributions from arbitrarily high masses can modify the numerator while leaving the normalization fixed.

To obtain a well-defined convex optimization problem, we therefore need a linear combination of sum rules whose high-energy kernel is positive semidefinite in all representation channels. We now investigate whether such combinations exist at successive subtraction orders.

At minus-one subtractions, the only nontrivial sum rule is the gluon sum rule \eqref{eq:gYMsumrule}. Since its kernel already contains both positive and negative entries, and there are neither additional sum rules nor null constraints available at this order, there is no possibility of constructing a positive combination. In particular, this means that we cannot normalize with respect to $g_{YM}^2$.

The situation improves at zero subtractions. Here we have the two forward-limit sum rules
\eqref{eq:forwardlevel0}, together with the null constraint $\mathcal{Y}^{(1)}_{0,0}$, \eqref{eq:topNCs} and the improved graviton sum rule \eqref{eq:gluonGsumrule}. Let us first ignore the graviton sum rule and consider only the forward-limit relations. Unfortunately, no linear combination of these matrices is positive semidefinite. A simple way to see this is that all available matrices are traceless, so any nontrivial combination necessarily contains at least one negative eigenvalue.

One is therefore forced to include the graviton sum rule. However, this introduces an additional subtlety. The graviton sum rule contains a $1/u$ pole and must eventually be smeared against suitable wavefunctions. Since smearing probes the impact parameter regime reviewed in App. \ref{app:wavepacket}, it is not sufficient to use the forward-limit expressions alone, one must instead work with the improved sum rules of App. \ref{app:extraSR}.

At the level of the unsmeared kernels, one can indeed find linear combinations of the four improved sum rules for which all six representation channels contribute positively in the limit $J,m \rightarrow\infty$ with $b=\frac{2J}{m}$ held fixed. One can furthermore impose positivity in the large-$m$, fixed-$J$ limit, and positive combinations continue to exist, such as
\begin{equation}
    -\frac{8\pi G}{u}+\frac{1}{2}g_0^{(1)}+\frac{3}{2}g_0^{(2)}=\left<\mathbb{D}(u)+\frac{1}{2}\mathbb{E}(u)+\frac{3}{2}\mathbb{G}(u)\right>\, ,
\end{equation}
where the diagonal matrices in the heavy average can be found in \eqref{eq:matrixD}, and \eqref{eq:matrixE} -- \eqref{eq:matrixG}. However, the allowed family is not unique, and after smearing all such combinations we examined develop negative regions at low mass. These regions moreover grow with increasing spin. Using the standard smearing wavefunctions, we were therefore unable to construct a fully positive normalization condition. While this does not constitute a proof of impossibility, it suggests that any successful implementation would likely require a more sophisticated treatment of the smearing problem. In particular, we are presently unable to use the graviton sum rule as a normalization condition either.

Proceeding to higher subtraction order, the first positive combination constructed purely from forward-limit sum rules and null constraints appears at two subtractions. One could therefore normalize with respect to such a combination. The drawback is that doing so effectively discards the low-subtraction sum rules, and hence much of the additional information gained from supersymmetry.

Instead, we adopt a different strategy and introduce an artificial normalization condition which does not correspond to an EFT coefficient of either amplitude. Concretely, we consider the positive-definite kernel
\begin{equation}
    C_k(m^2,J)=\frac{1}{m^{2k}}\,\mathbb{I}_{6\times6}\, ,
\end{equation}
where $\mathbb{I}_{6\times6}$ stands for the identity matrix. The remaining question is then to determine the smallest value of $k$ for which the corresponding high-energy average converges. To do so, we explicitly decompose the heterotic and Type I amplitudes into representation-channel residues and numerically evaluate the resulting sums over resonances. We find that convergence first occurs for $k = 1$:
\begin{equation}
    C_1=\left<\frac{1}{m^{2}}\,\mathbb{I}_{6\times6}\right>\, .
\end{equation}
In our conventions,
\begin{align}
    \begin{split}
        \text{Heterotic:}\qquad &\frac{C_1\,M^6}{g^2_{YM}}\approx38.9\, ,
        \\
        \text{Type I:}\qquad &\frac{C_1\,M^6}{g^2_{YM}}\approx50.6\, ,
    \end{split}
\end{align}
with $M^2=4/\alpha'$ and $M^2=2/\alpha'$ respectively.

This normalization condition allows us to formulate a conventional convex bootstrap problem starting at one subtraction. The price is that $C_1$ has mass dimension higher than $g_{YM}^2$, $G$, and the coefficients $g^{(i)}_0$. Consequently, these couplings cannot be bounded since one may always add spectral weight at arbitrarily large mass without affecting $C_1$. The normalization nevertheless suffices to constrain the coefficients $g^{(i)}_1$ and grants access to all but one null constraint. This is the strategy we will pursue in the next section.

\subsection{Basic bounds}

\begin{figure}[ht]
\centering
\includegraphics[width=0.8\textwidth]{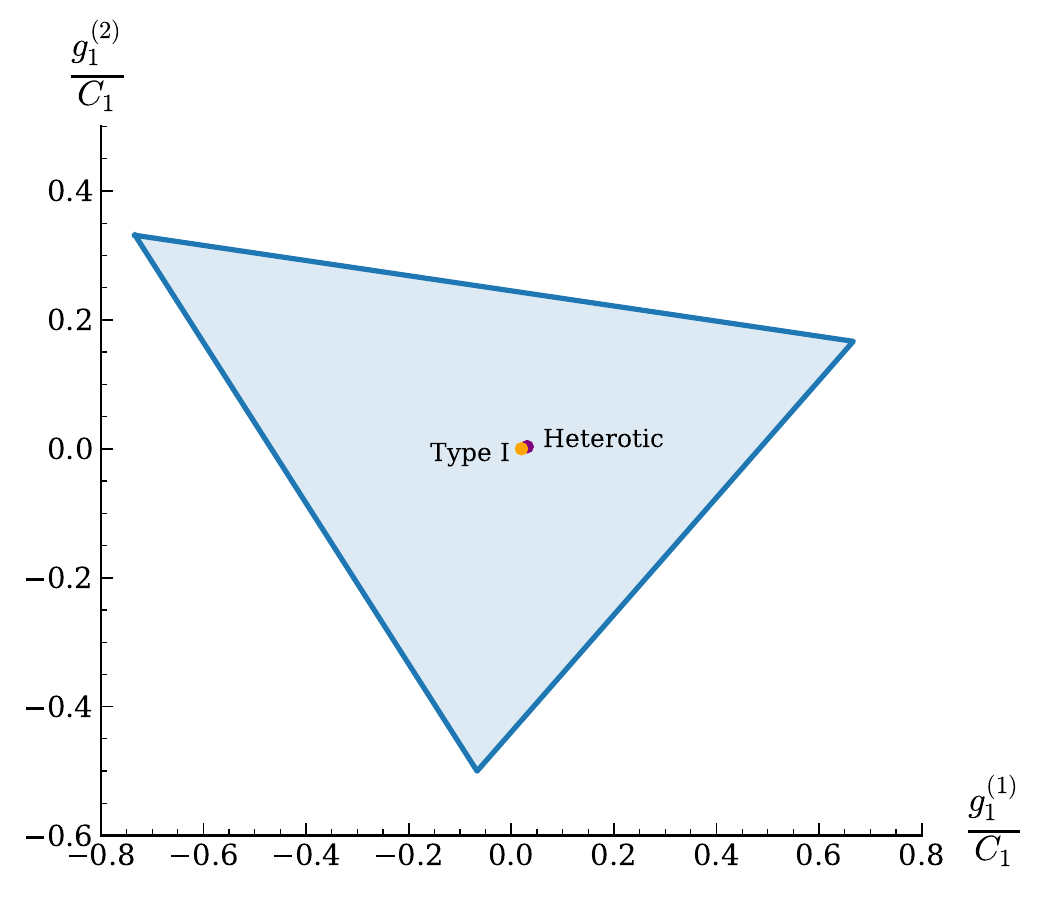}
\caption{Exclusion plot for  $g^{(1)}_1$ and $g^{(2)}_1$ normalized by $C_1$. The allowed region is shaded in blue. The plot is unaffected by the number of null constraints included. The values of the EFT coefficients for the two known consistent amplitudes, corresponding to the Type I (orange) and heterotic (purple) string theories, are shown on the plot.}
\label{fig:gluong1g1}
\end{figure}

As we have explained in the previous section, the usual procedure requires some modifications to be applicable in the present context. We will normalize with respect to the artificial normalization $C_1$, which makes the $g^{(i)}_1$ Wilson coefficients the first accessible ones. For the single trace part of the amplitude, the forward limit sum rules for this coupling reads
\begin{align}
    \label{eq:SingleTraceForwardSumRulesGluons}
    g^{(1)}_1 = \left< \frac{1}{15m^2}\:\text{diag}(0\,,1\,,-11\,,10\,,1\,,-1)\right>\, .
\end{align}
The corresponding sum rule for the double trace amplitude is
\begin{align}
    \label{eq:ForwardSumRulesGluons}
    g^{(2)}_1 = \left<\frac{1}{6 m^2}\:\text{diag}\!\left(-\frac{3}{248}\,,\frac{1}{40}\,,\frac{308}{155}\,,1\,,0\,,-3\right)\right>\, .
\end{align}
In addition to the sum rules, there are three null constraints at this level of subtraction. They are given by
\begin{align}\label{eq:OnceSubtractedNCs}
\begin{split}
    \mathcal{X}^{(1)}_{1,0}&= \frac{1}{m^2}\:\text{diag}\!\left(0\,,\mathcal{J}^2-4\,,9\mathcal{J}^2+4\,,-10\mathcal{J}^2,\mathcal{J}^2-4\,,4-\mathcal{J}^2\right)\, , \\
    \mathcal{Y}^{(1)}_{1,0}&= \frac{1}{m^2}\:\text{diag}\!\left(0\,,\mathcal{J}^2,20-\mathcal{J}^2,20\,,4-\mathcal{J}^2\,,\mathcal{J}^2-4\right)\, , \\
    \mathcal{X}^{(2)}_{1,0}&= \frac{1}{m^2}\:\text{diag}\!\left(\frac{3}{248}\mathcal{J}^2,-\frac{1}{40}\mathcal{J}^2,\frac{312}{155}\mathcal{J}^2-8\,,\mathcal{J}^2-4\,,0\,,3\mathcal{J}^2-12\right)\, ,
\end{split}
\end{align}
where $\mathcal{J}^2 = J(J+D-3)$ is the quadratic Casimir of $SO(D-1)$. One can obtain an exclusion plot for the allowed values of $g^{(1)}_1$ as $g^{(2)}_1$ is varied, normalizing by $C_1$. The resulting plot is shown in figure~\ref{fig:gluong1g1}. 

The exclusion plot is bounded by three extremal points, which can be understood analytically. Each is supported on finite-mass states in only a single representation channel and is located at
\begin{align}
\begin{split}
         \text{states in } R=3:\quad &\left(-\frac{11}{15}, \frac{154}{465}\right) \\[3pt]
         \text{states in } R=4:\quad &\quad\,\left(\frac{2}{3}, \frac{1}{6}\right) \\[3pt]
         \text{states in } R=6:\quad &\left(-\frac{1}{15}, -\frac{1}{2}\right)\, .
\end{split}
\end{align}
Up to states at infinite mass required to satisfy the null constraints, each extremal point is supported entirely in the indicated representation channel. The $R=3$ and $R=4$ solutions respectively minimize and maximize $g^{(1)}_1/C_1$, while the $R=6$ solution minimizes $g^{(2)}_1/C_1$. 

Both string theory amplitudes lie close to the origin, well within the allowed region. This is partly due to cancellations in the contributions of the different channels to the sum rules. Indeed, we have verified using the known heterotic spectrum that the sum rules are saturated by the contribution of the lowest lying massive states, and all contributions at higher masses cancel between the different channels. In introducing $C_1$, we limit this phenomenon, since no cancellations can occur in its sum rule, by virtue of it being positive definite. 

Note that the exclusion region is unaffected by null constraints; the same region can be obtained analytically by looking at the $C_1$ and $g^{(i)}_1$ sum rules. This can be understood as follows: all three relevant sum rules scales as $m^{-2}$, with no dependence on spin. The leading null constraint \eqref{eq:topNCs} is independent of $m$, so it can always be satisfied by adding states at infinite mass without affecting the relevant quantities. The null constraints coming in at $(1)$ subtraction, given in \eqref{eq:OnceSubtractedNCs} have terms scaling as $\frac{\mathcal{J}^2}{m^2}$ and $m^{-2}$. The contribution of states at finite mass can thus be canceled by placing states in the regime of fixed impact parameter $b=\frac{2J}{m}$ while taking $m\to\infty$. This doesn't affect the quantities being bounded since their sum rules are subleading in this regime. At higher subtractions, more null constraints come in, but they again scale as $b^k\times m^{-2\ell}$ for some $k,\ell \geq 0$. Crucially, they are always traceless, which ensures any contribution from finite mass states can be canceled by including states in appropriate infinite limits in parameter space. 

Qualitatively, the mechanism that trivializes the sum rules is that they are sign indefinite in the regime of fixed $b$ with $m\to\infty$, while the sum rules are subleading there. This again traces back to the matrices appearing in the sum rules failing to be sign definite. In more favorable cases, the sum rules have a certain sign at small spin, and opposite sign at large spin. In these cases, the null constraints genuinely constrain the high energy spectrum in terms of the low lying states, whereas here the null constraints can be trivialized by picking an appropriate high energy spectrum.

If we instead considered couplings of different mass dimension, we would obtain similar exclusion plots. The allowed region is again the convex hull of a finite set of extremal points, supplemented by limiting extremal configurations in which the coupling of highest mass dimension vanishes. These additional points arise by sending all finite-mass states to infinity and therefore correspond to spectra with parametrically large masses. The resulting boundaries are piecewise linear, connecting the various extremal points.

\section{Discussion and outlook}
\label{sec:discussion}

In this work we have extended the tree-level S-matrix bootstrap for quantum gravity to ten-dimensional theories with half-maximal supersymmetry. The half-maximal setting is rich enough to include the heterotic string, and at the same time sufficiently constrained to allow a systematic dispersive analysis of four-point amplitudes. Our results show that several of the qualitative features found in the maximally supersymmetric bootstrap persist with less supersymmetry. In particular, in the gravitational sector much of the boundary of the allowed region is again controlled by extremal amplitudes whose spectra organize into a single linear Regge trajectory, or by convex combinations of such amplitudes. In the gluon sector, treated without taking a planar limit, the analysis exposes a new structural complication: the positivity basis is naturally organized by irreducible representation channels, while the low-energy EFT data are organized by single- and double-trace structures.

This last point is perhaps the main limitation of the present analysis. In the purely gluonic problem, the sum rules associated with the Yang--Mills coupling and Newton's constant are not positive in the representation-channel basis. As a result, neither $g_{\rm YM}$ nor $G$ provides a natural normalization for the convex bootstrap problem. We have therefore used an auxiliary positive normalization to obtain preliminary bounds. These bounds are meaningful, but they do not yet isolate the physically most interesting dimensionless ratio controlling the relative strength of gauge and gravitational interactions. 

A natural next step is to study the coupled system of gluon, graviton and mixed gluon--graviton amplitudes. Mixed amplitudes contain both gauge and gravitational factorization channels, and the same three-point couplings enter different amplitudes in correlated ways. One may therefore hope that the coupled system supplies positivity and factorization constraints that are invisible in the gluon sector alone. In particular, such a bootstrap may provide a more direct way of constraining the ratio of gravitational to gauge interactions in heterotic-like theories.

Another important direction is to incorporate higher-point amplitudes. Four-point dispersive bounds are powerful, but they are linear in the spectral densities and therefore naturally lead to convex optimization problems. Higher-point factorization imposes additional nonlinear constraints on the same low-energy data. Recent work in maximally supersymmetric gauge theory has shown that such constraints can be strong enough to push the allowed EFT data toward the open-string amplitude~\cite{Elvang:2026pmc}. It would be very interesting to understand whether analogous higher-point constraints in the half-maximal, non-planar setting can further distinguish the heterotic string from the more economical extremal solutions that saturate the four-point bounds.

More broadly, the results of this paper reinforce the picture that emerged in the maximally supersymmetric analysis. The basic principles of analyticity, unitarity, crossing symmetry, supersymmetry and Regge boundedness already impose highly structured constraints on weakly coupled theories of gravity. At the boundaries of the allowed regions one repeatedly encounters minimal Regge-like solutions: single linear trajectories, shifted intercepts, variable slopes, and limiting infinite-spin configurations. It is a curious fact  that in both the maximal and half-maximal supersymmetric cases, and both for gravitons and gluons, extremal amplitudes often exhibit single {\it linear} trajectories, while in the large $N$ pion bootstrap~\cite{Albert:2022oes} the extremal amplitude comprises a single {\it curved} trajectory~\cite{Albert:2023seb}. Linear trajectories saturate the sequential spin constraints of~\cite{Berman:2024owc}.

Known string amplitudes obey the bounds but need not sit at the boundary. The challenge, therefore, is not simply to show that string theory is allowed, but to identify which additional consistency conditions select full string amplitudes from the larger space of amplitudes compatible with four-point tree-level consistency. The half-maximal bootstrap studied here provides a useful step in this direction.

\section*{Acknowledgements}
We are grateful to Jan Albert, Deniz Bozkurt, Sebastian Harris, Yue-Zhou Li, and Torben Skrzypek for interesting discussions and comments. MB is supported by the FRQNT doctoral scholarship.
WK is supported by ERC-2021-CoG - BrokenSymmetries 101044226. LR is partially supported by the NSF grant PHY-2513893 and by the Simons Foundation grant 681267 (Simons Investigator Award).

The authors would like to thank Stony Brook Research Computing and Cyberinfrastructure and the Institute for Advanced Computational Science at Stony Brook University for access to the high-performance SeaWulf computing system, which was made possible by \$1.85M in grants from the National Science Foundation (awards 1531492 and 2215987) and matching funds from the the Empire State Development’s Division of Science, Technology and Innovation (NYSTAR) program (contract C210148).

\appendix

\section{Playing with wavefunctions}\label{app:wavepacket}

As discussed in section \ref{sec:gravitybounds}, convergence of the numerical bounds becomes noticeably slower near the $\tilde g_2=0$ axis. The $g_2$ sum rule \eqref{eq:g2SumRule} implies that approaching this regime requires pushing all resonances to parametrically large mass, $m\to\infty$. This motivates modifying the wavepacket basis in order to improve sensitivity in the large-mass region.

Most of the discussion below is a brief review of the impact parameter analysis of smeared sum rules, developed in \cite{Caron-Huot:2021rmr}. The only new ingredient is the inclusion of a wavepacket localized at the cutoff scale, which may be viewed as a limiting case of the power-law basis considered there.

In the main text, see \eqref{eq:Gsmeared}, we considered wavepackets of the form
\begin{equation}
    f_k(p)=p^{k+\frac12}\, ,
\end{equation}
where $u=-p^2$. Here we supplement this basis with a wavepacket localized at the cutoff scale,
\begin{equation}
    f_\infty(p)=\delta(p-M),
\end{equation}
which may heuristically be viewed as the $k\to\infty$ limit of the power-law basis functions.

In the limit $m\to\infty$ with $b=\frac{2J}{m}$ held fixed, the Gegenbauer polynomial appearing in the smeared $G$--$g_0$ sum rule \eqref{eq:Gsmeared} reduces to a Bessel function. More precisely, the right-hand side becomes
\begin{equation}
    \Gamma\!\left(\frac{D-2}{2}\right) \int_0^M dp\, f(p)\, \frac{J_{\frac{D-4}{2}}(bp)} {(bp/2)^{\frac{D-4}{2}}}\, ,
\end{equation}
where $J_\nu$ denotes the Bessel function of the first kind. For the power-law wavepackets $f_k(p)$, this integral can be evaluated analytically:
\begin{equation} 
    \Gamma\!\left(\frac{D-2}{2}\right) \int_0^M dp\, p^n \frac{J_{\frac{D-4}{2}}(bp)} {(bp/2)^{\frac{D-4}{2}}} = M^{n-\frac{D-2}{2}}\frac{\, _1F_2\left(\frac{n+1}{2};\frac{D-2}{2},\frac{n+3}{2};-\frac{b^2M^2}{4}\right)}{n+1}\, . 
\end{equation}
As discussed in \cite{Caron-Huot:2021rmr}, the large-$b$ expansion of this takes the schematic form
\begin{equation}
    A(b)+B(b)\cos\!\left(bM-\frac{\pi(D-1)}{4}\right)+C(b)\sin\!\left(bM-\frac{\pi(D-1)}{4}\right)\, ,
\end{equation}
where $A(b)$, $B(b)$, and $C(b)$ admit asymptotic expansions in inverse powers of $b$. Positivity can then be imposed through the stronger positive semidefinite condition
\begin{equation}
    \begin{pmatrix}
        A(b)+B(b) & C(b) \\
        C(b) & A(b)-B(b)
    \end{pmatrix}
    \succeq 0\, .
\end{equation}
At large $b$, positivity requires the non-oscillatory contribution $A(b)$ to dominate over the oscillatory terms. For the power-law wavepackets, the leading non-oscillatory contribution behaves as
\begin{equation}
    A(b)\sim b^{-(k+\frac{3}{2})}\, ,
\end{equation}
while the oscillatory terms scale universally as
\begin{equation}
    B(b)\sim b^{-\frac{D-1}{2}} \quad \text{and} \quad C(b)\sim b^{-\frac{D+1}{2}}\, .
\end{equation}
Consequently, positivity at large $b$ requires the basis to contain sufficiently low-power wavepackets such that
\begin{equation}
    k \leq \frac{D}{2}-2\, .
\end{equation}
For the localized wavepacket $f_\infty$, the non-oscillatory contribution vanishes identically
\begin{equation}
    A(b)=0\, ,
\end{equation}
so the large-$b$ behavior is purely oscillatory, with
\begin{equation}
    B(b)\sim b^{-\frac{D-3}{2}} \quad \text{and} \quad C(b)\sim b^{-\frac{D-1}{2}}\, .
\end{equation}
As a result, positivity cannot be maintained using $f_\infty$ alone. This is simply the most extreme manifestation of the general fact discussed above: sufficiently high-power wavepackets must always be accompanied by lower-power wavepackets whose non-oscillatory contributions dominate at large impact parameter.\footnote{This also clarifies why one cannot improve the Regge behavior by simply constructing sum rules around finite $u$. In other $S$-matrix bootstrap settings, it was hoped that continuing the Regge trajectory to finite $u$ could strengthen the asymptotic behavior. However, this effectively corresponds to a localized wavepacket of the type discussed here, which by itself cannot satisfy positivity at large impact parameter.}

\begin{figure}[ht]
    \centering
    \includegraphics[width=.9\textwidth]{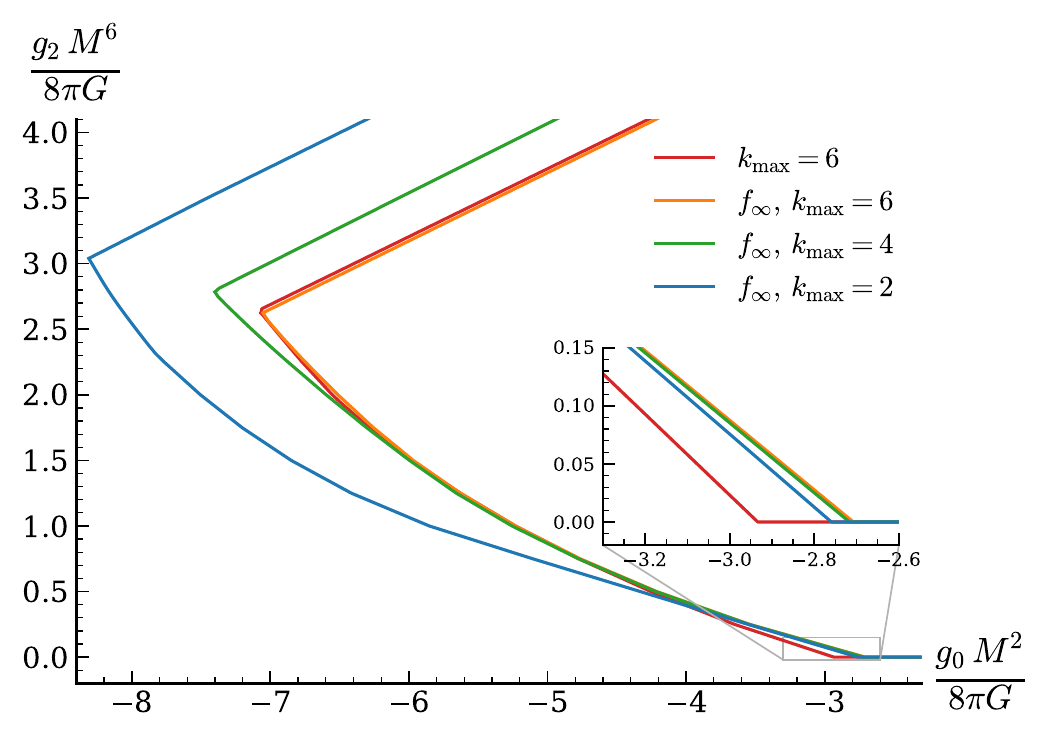}
    \caption{Exclusion plot in the $(\tilde g_0,\tilde g_2)$ plane illustrating the effect of supplementing the power-law wavepacket basis with the localized wavepacket $f_\infty$. All bounds are obtained with $n_{\mathrm{max}}=13$. The red curve uses only the power-law basis with $k_{\mathrm{max}}=6$, while the remaining curves additionally include $f_\infty$ together with different values of $k_{\mathrm{max}}$. Near the $\tilde g_2=0$ axis, the bounds obtained using the enlarged basis converge significantly faster and already outperform the pure power-law basis even when fewer wavepackets are included.}
    \label{fig:deltaAdded}
\end{figure}

Including the additional localized wavepacket improves convergence near $\tilde g_2=0$, as illustrated in figure \ref{fig:deltaAdded}. Even at relatively low truncation order, the resulting lower bounds on $\tilde g_0$ improve noticeably compared to the basis without the additional wavepacket.

Despite this improvement, we do not systematically employ the enlarged basis in the main numerical analysis. At nonzero $\tilde g_2$, the effect on the bounds is no better than replacing $f_\infty$ by the next higher $f_k$, while the additional wavepacket complicates the large-spin and small-mass approximations used to increase the cutoff in the numerical implementation.

\section{More sum rules for gluon scattering}\label{app:extraSR}

Instead of working directly in the forward limit, one can improve the sum rules by subtracting higher-order forward-limit relations so as to isolate individual EFT couplings. This procedure is completely analogous to the construction of the improved graviton sum rule in the main text and leads to the following $u$-dependent sum rules for the two lowest EFT coefficients:

\begin{equation}
    \begin{alignedat}{2}
        \mathcal{U}^{(1)}_{0}-\sum^\infty_{k=1} u^k\, \mathcal{U}^{(1)}_k(u=0): &&\qquad g_0^{(1)} = \langle \mathbb{E}(u)\rangle\, , 
        \\
        \mathcal{V}^{(1)}_0 -\sum^\infty_{k=1} (-u)^k\, \mathcal{U}^{(1)}_k(u=0): &&\qquad g^{(1)}_0 = \langle \mathbb{F}(u)\rangle\, , 
        \\
        \mathcal{U}^{(2)}_{0}-\sum^\infty_{k=1} u^k\, \mathcal{U}^{(2)}_k(u=0): &&\qquad g_0^{(2)} = \langle \mathbb{G}(u)\rangle\, ,
    \end{alignedat}
\end{equation}
where the matrices $\mathbb{E}(u)$, $\mathbb{F}(u)$, and $\mathbb{G}(u)$ are diagonal. Their non-vanishing entries are given by
\begin{align}
    \begin{split}\label{eq:matrixE}
        \mathbb{E}_{11} &= 0 
        \\
        \mathbb{E}_{22} &=\frac{1}{15}\left(\mathcal{P}_J\!\left(\mbox{$1+\frac{2u}{m^2}$}\right)-\frac{u}{m^2-u}\right)
        \\
        \mathbb{E}_{33} &=\frac{1}{15}\left(\frac{9m^2-u}{m^2+u}\mathcal{P}_J\!\left(\mbox{$1+\frac{2u}{m^2}$}\right)+\frac{u\,(11m^2-9u)}{m^4-u^2}\right)
        \\
        \mathbb{E}_{44} &=-\frac{2}{3}\left(\frac{m^2}{m^2+u}\mathcal{P}_J\!\left(\mbox{$1+\frac{2u}{m^2}$}\right)+\frac{u}{m^2+u}\right)
        \\
        \mathbb{E}_{55} &=\frac{1}{15}\left(\mathcal{P}_J\!\left(\mbox{$1+\frac{2u}{m^2}$}\right)-\frac{u}{m^2-u}\right)
        \\
        \mathbb{E}_{66} &=-\frac{1}{15}\left(\mathcal{P}_J\!\left(\mbox{$1+\frac{2u}{m^2}$}\right)-\frac{u}{m^2-u}\right)
    \end{split}
\\[1.4em]
    \begin{split}\label{eq:matrixF}
        \mathbb{F}_{11} &= 0
        \\
        \mathbb{F}_{22} &=\frac{1}{15}\left(\frac{2m^2+u}{m^2+u}\mathcal{P}_J\!\left(\mbox{$1+\frac{2u}{m^2}$}\right)+\frac{u}{m^2+u}\right)
        \\
        \mathbb{F}_{33} &=-\frac{1}{15}\left(\frac{2m^2+u}{m^2+u}\mathcal{P}_J\!\left(\mbox{$1+\frac{2u}{m^2}$}\right)+\frac{u\,(11m^2+9u)}{m^4-u^2}\right)
        \\
        \mathbb{F}_{44} &=\frac{2u}{3(m^2-u)}
        \\
        \mathbb{F}_{55} &=-\frac{1}{15}\left(\frac{2m^2+u}{m^2+u}\mathcal{P}_J\!\left(\mbox{$1+\frac{2u}{m^2}$}\right)-\frac{u}{m^2+u}\right)
        \\
        \mathbb{F}_{66} &=\frac{1}{15}\left(\frac{2m^2+u}{m^2+u}\mathcal{P}_J\!\left(\mbox{$1+\frac{2u}{m^2}$}\right)-\frac{u}{m^2+u}\right)
    \end{split}
\end{align} 
\begin{align}
    \begin{split}\label{eq:matrixG}
        \mathbb{G}_{11} &=\frac{1}{496}\left(\frac{m^2}{m^2+u}\mathcal{P}_J\!\left(\mbox{$1+\frac{2u}{m^2}$}\right)+\frac{u}{m^2+u}\right)
        \\
        \mathbb{G}_{22} &=-\frac{1}{240}\left(\frac{m^2}{m^2+u}\mathcal{P}_J\!\left(\mbox{$1+\frac{2u}{m^2}$}\right)+\frac{u}{m^2+u}\right)
        \\
        \mathbb{G}_{33} &=\frac{1}{465}\left(\frac{156m^2+155u}{m^2+u}\mathcal{P}_J\!\left(\mbox{$1+\frac{2u}{m^2}$}\right)-\frac{2u\,(77m^2+78u)}{m^4-u^2}\right)
        \\
        \mathbb{G}_{44} &=\frac{1}{6}\left(\mathcal{P}_J\!\left(\mbox{$1+\frac{2u}{m^2}$}\right)-\frac{u}{m^2-u}\right)
        \\
        \mathbb{G}_{55} &= 0
        \\
        \mathbb{G}_{66} &=-\frac{1}{2}\left(\mathcal{P}_J\!\left(\mbox{$1+\frac{2u}{m^2}$}\right)-\frac{u}{m^2-u}\right)\, .
    \end{split}
\end{align}
Taking the forward limit $u\to 0$ reproduces the sum rules in \eqref{eq:forwardlevel0}.

\bibliographystyle{jhep}
\bibliography{references}

\end{document}